\documentclass[11pt]{article}

\usepackage[T1]{fontenc}
\usepackage[cp1251]{inputenc}
\usepackage{textcomp}
\usepackage[centertags]{amsmath}
\usepackage[mediummath]{nccmath}
\usepackage{amsfonts}
\usepackage{amssymb}
\usepackage{braket}
\usepackage[pdftex]{hyperref}
\usepackage{graphicx}
\usepackage[numbers,sort&compress]{natbib}
\usepackage{comment}

\usepackage{paperinitial}

\paperinitialization{15mm}{15mm}{15mm}{15mm}{2pt}{10pt}

\DeclareMathOperator{\re}{Re}
\DeclareMathOperator{\im}{Im}
\DeclareMathOperator{\sgn}{sgn}

\DeclareMathOperator{\Sp}{Sp}

\newcommand{\lan}{\langle}
\newcommand{\ran}{\rangle}
\newcommand{\bs}{\boldsymbol}

\newcommand{\e}{\varepsilon}
\newcommand{\vf}{\varphi}
\newcommand{\s}{\sigma}
\newcommand{\Si}{\Sigma}
\newcommand{\al}{\alpha}
\newcommand{\be}{\beta}
\newcommand{\ga}{\gamma}
\newcommand{\Ga}{\Gamma}
\newcommand{\de}{\delta}
\newcommand{\De}{\Delta}
\newcommand{\vk}{\varkappa}
\newcommand{\la}{\lambda}

\newcommand{\ups}{\upsilon}

\newcommand{\spx}{\mathbf{x}}

\newcommand{\spp}{\mathbf{p}}
\newcommand{\spq}{\mathbf{q}}
\newcommand{\spk}{\mathbf{k}}
\newcommand{\spg}{\mathbf{g}}

\newcommand{\spb}{\mathbf{b}}

\newcommand{\spn}{\mathbf{n}}
\newcommand{\spd}{\mathbf{d}}

\begin{document}
\allowdisplaybreaks[4]
\frenchspacing
\setlength{\unitlength}{1pt}
\title{{\Large\textbf{Coherent elastic scattering of low energy photons by neutrons}}}

\date{}
	
\author{
	P.O. Kazinski${}^{1)}$\thanks{E-mail: \texttt{kpo@phys.tsu.ru}}\;
	and A.A. Sokolov${}^{1),2)}$\thanks{E-mail: \texttt{asokolov@tpu.ru}}\\[0.5em]
	{\normalsize ${}^{1)}$ Physics Faculty, Tomsk State University, Tomsk 634050, Russia}\\
	{\normalsize ${}^{2)}$ Tomsk Polytechnic University, Tomsk 634050, Russia}
}
	
\maketitle

\begin{abstract}

The Compton process with the initial states of photons and neutrons described by the density matrices of a general form is studied for low energies of photons. The coherent contribution to the inclusive probability to record a photon is investigated in detail. This contribution gives the hologram of the neutron one-particle density matrix. The evolution of the Stokes parameters of scattered photons is described. The susceptibility tensor of a neutron gas and a wave packet of a single neutron is obtained. The explicit expression for the photon polarization operator in the presence of free neutrons is derived. It turns out that this polarization operator possesses pole singularities in the short wavelength approximation. These singularities corresponding to the additional degrees of freedom are identified with plasmons and the respective plasmon-polaritons are described. There are eight independent plasmon-polariton modes in a neutron gas and on a single neutron wave packet. Some plasmon-polariton modes prove to be tachyonic and unstable manifesting a spontaneous generation of the magnetic field. The estimates of the parameters of the neutron gas when it becomes ferromagnetic are found. In the infrared limit, the neutron wave packet behaves in coherent Compton scattering as a point particle with dynamical magnetic moment, the additional degrees of freedom being reduced to the dynamical part of the magnetic moment.

\end{abstract}
	
\section{Introduction}

At present, classical holography based on photons, electrons, and neutrons is the standard tool for investigation of properties of objects of various nature \cite{Gabor1949,Cser2004,SpBookMicrosc2019,Winkler2020,Harada2021,Inoue2022,Daimon2024}. It was shown in the recent papers \cite{pra103,KazSol2022,KazSol2023,KRS2023,radet,AKS2025} that a similar technique can be employed to probe the wave functions of particles, including their dynamics and collapse, by constructing the holograms. Such holograms arise naturally in coherent scattering processes \cite{pra103,KazSol2022,KazSol2023,KRS2023,radet,AKS2025,BednNaum2021,KazinskiFr24} where the initial state of a target particle or a collective of target particles does not change and coincides with the final state in the interaction picture. It is relevant in these processes that the initial state of a probe particle is not a plane wave. In the plane-wave approximation, the terms in the $S$-matrix responsible for the coherent contributions are different from zero only on the measure zero subset of the momentum space and are commonly discarded \cite{SchwartzBook2014}. On squaring the $S$-matrix, these contributions result in the interference terms between the connected part of the $S$-matrix and the terms coming from the through lines of the Feynman diagrams or, more generally, between the different connected parts of the $S$-matrix (see for details \cite{KazinskiFr24}). In particular, the differential cross section does not contain the coherent contributions by construction (see, e.g., \cite{WeinbergB.12}) and in this sense is determined by the incoherent contribution to the probability to record the probe particle.

It turns out \cite{KazSol2022,KRS2023} that, in scattering of the probe particle wave packet by the quantum object (the target), the coherent contribution to the probability to record the probe particle, i.e., the hologram, is the same as if scattering were occurring on a fluid consisting of such objects with a density equal to the probability density constructed from the wave function of the target. As for the incoherent contribution to record the probe particle, in particular, for the differential cross section, this is not the case \cite{MarcuseI,PMHK08,Corson2011,CorsPeat11,LMKMPY23,Wong21,KdGdAR21,Remez19,PanGov18,WCCP16}. In a certain approximation, the incoherent contribution is such as if scattering were occurring on a collective of target objects, the concentration of which is determined by the wave function of the target but the scattering probabilities are added up and not the amplitudes \cite{PMHK08,Remez19,radet}. The analogy of coherent scattering with scattering by macroscopic objects allows us to introduce the macroscopic characteristics, such as the dielectric susceptibility tensor, for microscopic objects including elementary particles. In the papers \cite{KazSol2022,KazSol2023,AKS2025}, such tensors were constructed for the wave packets of a single electron and a single photon and, in the paper \cite{AKS2025}, the polarization operator of a photon in the presence of a wave packet of an electron was found out of the photon  mass-shell. The examination of singularities of the polarization operator and the respective effective Maxwell equations revealed that there are quasiparticles -- the plasmons and the corresponding plasmon-polaritons -- even on the wave packet of a single electron.

In the present paper, we carry out an analogous analysis for the hologram of a quantum state of a neutron gas and a wave packet of a single neutron. This hologram corresponds to the coherent contribution to the probability to record a low energy photon in elastic scattering by neutrons. Such a contribution is of lower order in the coupling constant, i.e., it gives the leading contribution, in comparison with the ``standard'' incoherent contribution entering into the differential cross section. We derive the general expression for the hologram and investigate thoroughly its properties in the case of small momentum transfer from the photon to the neutron for both the unpolarized and polarized particles. In accordance with the theory developed in \cite{KazSol2022,KazSol2023,AKS2025}, these scattering data can be interpreted as scattering by a fluid with a certain dielectric susceptibility tensor. We obtain the general expression for such a tensor and consider some particular cases where this expression is simplified and admits a clear physical interpretation. In particular, we show that the neutron prepared in a spin polarized quantum state participates in coherent Compton scattering with a low energy photon as a gyrotropic medium. As is known \cite{KLein1955,Baldin1960,Petrunkin1968,Baranov1976,Grieshammer2012}, in describing elastic scattering of low energy photons by hadrons, the phenomenological parameters arise that characterize the electric and magnetic polarizabilities of hadrons. These parameters have nothing in common with the dielectric susceptibility tensor that we study in the present paper although these parameters do influence the dielectric susceptibility tensor for sufficiently large energies of probe photons.

In addition to the study of coherent scattering of photons by neutrons, we deduce the one-loop photon polarization operator out of the photon mass-shell. The mass-shell expression for this operator coincides, up to a common factor, with the dielectric susceptibility tensor. The procedure for derivation of the expression for this tensor is based on the $in$-$in$ perturbation theory \cite{Schw1961,Keld64,CSHY85,GFSh.3,DeWGAQFT.11,CalzHu} and is presented in detail for a gas of electrons in \cite{AKS2025}. Therefore, in the present paper, we immediately give the general expression for this polarization operator and scrutinize it in several particular cases. The analysis reveals that there are the additional degrees of freedom -- quasiparticles (the plasmons) -- in a neutron gas and on a wave packet of a single neutron. Hybridization of these plasmons with the electromagnetic field gives rise to plasmon-polaritons. As in the case of a single electron wave packet, there are eight independent plasmon-polariton modes on a single neutron wave packet. For a neutron gas, some of these plasmon-polariton modes are unstable and possess a tachyonic dispersion law in certain regions of momenta. This instability is interpreted as the onset of spontaneous magnetization of the neutron gas at sufficiently low temperatures and sufficiently hight densities, i.e., in this domain of parameters the neutron gas is ferromagnetic. For example, the nondegenerate gas of ultracold neutrons \cite{Ignatovich1996,Serebrov2011,Pokotilovski2018,Lauss2021} at a temperature of $1.06$ neV goes to the ferromagnetic state for the concentrations of order $0.5$ mol/cm${}^{3}$ and larger. This concentration is by fifteen orders smaller than the nuclear concentration. Notice that, for the degenerate gas of free neutrons with density of order of the nuclear density, the ferromagnetic instability is absent \cite{Delsante1979,Anand1981}. However, there are the models \cite{Akhiezer1996} where the nuclear interaction of neutrons changes the behavior of a neutron gas at nuclear densities and makes it ferromagnetic.

Besides, we investigate the infrared limit of the photon polarization operator in the presence of spin polarized neutrons and the corresponding effective Maxwell equations. In that case, coherent Compton scattering proceeds as if the neutron were a neutral point particle with dynamical magnetic moment. Then the additional degrees of freedom boil down to this dynamical magnetic moment.

The paper is organized as follows. We formulate in Sec. \ref{Model_and_Notation} the model at issue and the notation adopted. In Sec. \ref{Inclusive_Probability}, we derive the general formula for the inclusive probability to record a photon in coherent Compton scattering, i.e., the general formula for the hologram of a neutron state. In Sec. \ref{Density_Matrix_of_Scattered_Photons}, the spin density matrix of scattered photons is analyzed in the leading nontrivial order of perturbation theory. The evolution of the Stokes parameters of coherently scattered photons is described in Sec. \ref{Evolution_of_Stokes_Parameters}. In Sec. \ref{Gaussian_Wave_Packets}, a change of these parameters is explicitly found for the Gaussian one-particle density matrices of photons and neutrons. Section \ref{Dielectric_Suscept_of_Neutr} is devoted to the dielectric susceptibility of a neutron wave packet and a neutron gas. The on- and off-shell expressions for the photon polarization operator in the presence of a single neutron or a neutron gas are derived there. The properties of the photon polarization operator and the corresponding effective Maxwell equations are also investigated in this section. In Conclusion, we summarize the results and discuss the prospects for further research. Some huge formulas and calculations are removed to appendices.

We follow the notation adopted in \cite{KazSol2022,KRS2023}. The Greek indices $\mu$, $\nu$, $\ldots$ are the space-time indices taking the values $\overline{0,3}$ and the Latin $i$, $j$ are the spatial indices. The summation over repeated indices is always understood unless otherwise stated. In certain cases, we will write this summation explicitly. We also suppose that the quantum states of particles are normalized to unity in some sufficiently large volume $V$. We use the system of units such that $\hbar=c=1$ and $e^2=4\pi\al$, where $\al$ is the fine structure constant. The Minkowski metric $\eta_{\mu\nu}$ is taken with the mostly minus signature. We follow the agreements
\begin{equation}
    \s^{\mu\nu}=\frac{i}{2}[\ga^\mu,\ga^\nu],\qquad \ga^5=-i\ga^0\ga^1\ga^2\ga^3,\qquad \e^{0123}=1,\qquad \e_{123}=1,
\end{equation}
and always use the standard representation for the $\ga$-matrices. The bar over Dirac spinors means the Dirac conjugation whereas this bar over other quantities denotes the complex conjugation.

\section{Model and notation}\label{Model_and_Notation}

The elastic scattering amplitude of photons by spin $1/2$ hadrons is described phenomenologically by the six form factors \cite{Baranov1976,Prange1958,Hearn1962,Bardeen1968,Grieshammer2012} which are usually taken from experiments. For the photon energies much lower than the rest energy of $\pi$-meson, these form factors can be expanded in the small photon energy that results in the expression with a few parameters of the hadron: the charge, the anomalous magnetic moment, and the electric and magnetic polarizabilities  \cite{KLein1955,Baldin1960,Petrunkin1968,Baranov1976,Grieshammer2012}. In the present paper, we consider even lower photon energies where the contributions of these electric and magnetic polarizabilities can be neglected and the Compton amplitude for a particle of an arbitrary spin is determined only by the particle charge and its magnetic moment \cite{WeinbergB.12,Low1954,GellMann1954,Weinberg1970,Grieshammer2012}. This simplifies calculations and underlines the fact that the hadron polarizabilities are not related to the susceptibility and polarization tensors discussed in the present paper. Furthermore, as a rule, the coherent effects stemming from the form of the particle wave packets are more likely to be seen for low energy photons. Thus, in the low energy limit, the Lagrangian density for hadrons interacting only electromagnetically and elastically with photons reads (see, e.g., \cite{LandLifQED})
\begin{equation}\label{QED_hadron_electron}
	\mathcal{L} = -\frac14 F_{\mu\nu}F^{\mu\nu}+\bar{\psi} (i\ga^\mu\partial_\mu-M)\psi -\bar{\psi} \Ga^\mu A_\mu\psi,
\end{equation}
where $M$ is the hadron mass, $F_{\mu\nu}:=\partial_{[\mu} A_{\nu]}$ is the electromagnetic stress tensor, $\psi$ is the Dirac spinor describing the quantum hadron field of spin $1/2$, $\Ga^\mu$ is an operator of electromagnetic interaction of a hadron, which has the form in the momentum representation \cite{LandLifQED}
\begin{equation}\label{hadron_curr}
	\bar{u}^{s'}(\spp')\Ga^\mu(p',p) u^s(\spp)=\bar{u}^{s'}(\spp')(e_p\ga^\mu+\mu_p i\sigma^{\mu\nu}k_\nu) u^s(\spp),
\end{equation}
where $e_p$ is the particle charge, $\mu_p$ is its anomalous magnetic moment, and the momentum $k^\mu$ inflows into the interaction vertex. Recall that $\mu_p=1.79 |e|/(2M)$ for a proton and $\mu_p=-1.91 |e|/(2M)$ for a neutron. The action \eqref{QED_hadron_electron} must be gauge fixed. Henceforth, we imply the Feynman gauge.

We adopt the normalization of the Dirac spinors and the mode functions of the electromagnetic field as in \cite{BaKaStrbook,pra103}. In particular,
\begin{equation}
	\begin{split}
		\hat\psi(x)&=\sum_s\int \frac{V d\spp}{(2\pi)^3} \sqrt{\frac{M}{V p_0}} \Bigl[u^s(\spp)e^{-ip_\mu x^\mu}
		\hat{a}_{s}(\spp)+v^s(\spp)e^{i p_\mu x^\mu} \hat{b}_{s}^{\dag}(\spp)\Bigr],
	\end{split}
\end{equation}
where the mass-shell condition $p_0=\sqrt{M^2+\spp^2}$ is fulfilled, $\hat{a}^\dag_{s}(\spp)$, $\hat{a}_{s}(\spp)$ are the creation and annihilation operators of hadrons $h$, respectively. The creation and annihilation operators denoted by $\hat{b}^\dag$, $\hat{b}$ are for the corresponding antiparticles $h^c$. Besides,
\begin{equation}\label{spinors}
		u^s(\spp)=\frac{M+\hat p}{\sqrt{2M(p_0+M)}} \genfrac{[}{]}{0pt}{}{\chi_s}{0},\qquad \ups^s(\spp)
		=\frac{M-\hat p}{\sqrt{2M(p_0+M)}} \genfrac{[}{]}{0pt}{}{0}{\chi_s},
\end{equation}
where the hat over a $4$-vector means the contraction of this $4$-vector with $\ga_\mu$ and
\begin{equation}
	(\bs\tau \bs\s) \chi_s=s \chi_s,\qquad s=\pm1.
\end{equation}
The unit vector,
\begin{equation}
	\bs\tau=(\sin\theta\cos\vf,\sin\theta\sin\vf,\cos\theta),
\end{equation}
specifies the direction of the spin projection quantization axis. There are the relations
\begin{equation}\label{energy_proj}
\begin{aligned}
	\sum_s u^s(\spp)\bar u^s(\spp)&=\frac{M+\hat p}{2M},&\qquad\sum_s \ups^s(\spp)\bar \ups^s(\spp)&=-\frac{M-\hat p}{2M},\\
	u^\pm(\spp)\bar u^\pm(\spp)&=\frac{M+\hat p}{2M}\frac{1\mp\ga^5\hat s}{2},&\qquad \ups^\pm(\spp)\bar \ups^\pm(\spp)&=
    -\frac{M-\hat p}{2M}\frac{1\pm\ga^5\hat s}{2},
\end{aligned}
\end{equation}
where the spin vector
\begin{equation}\label{s_mu_defn}
	s^\mu:=\Big(\frac{(\boldsymbol{\tau}\spp)}{M},\boldsymbol{\tau}+\frac{\spp(\boldsymbol{\tau}\spp)}{M(p_0+M)}\Big),
    \qquad s^\mu p_\mu=0,\qquad s^2=-\boldsymbol{\tau}^2.
\end{equation}
Furthermore, we use the condensed indices and notation. For example,
\begin{equation}
	\al:=(s,\spp),\qquad \sum_\al:=\sum_s \int \frac{Vd\spp}{(2\pi)^3},
\end{equation}
and $\al':=(s',\spp')$.

\section{Inclusive probability}\label{Inclusive_Probability}

Let us assume that the photons and the hadrons are not correlated at the initial instant of time $t_{in}\rightarrow-\infty$. The density matrix of such a system becomes
\begin{equation}
	\hat R=\hat R_{ph}\otimes \hat R_h\otimes |0\rangle_{h^c}\langle0|_{h^c},
\end{equation}
where $\hat R_{ph}$ is the photon density matrix, $\hat R_h$ is the hadron density matrix, and $|0\rangle_{h^c}$ is the vacuum state for the corresponding antiparticles. Hereinafter it is understood that the states and the operators are written in the interaction representation. For definiteness, we suppose that the photons were initially in the coherent state with the density matrix
\begin{equation}\label{R_ph}
	\hat R_{ph}=|d\rangle\langle \bar{d}|e^{-\bar d_\ga d_\ga},\qquad \hat{c}_{\ga'}|d\ran =d_{\ga'} |d\ran,
    \qquad |d\ran=e^{\hat{c}^\dag_\ga d_\ga}|0\ran,
\end{equation}
where $d_\ga$ is the complex amplitude of coherent state at the instant of time $t=0$, $\hat c^\dag_\ga$, $\hat c_\ga$ are the creation-annihilation operators for a photon with quantum numbers $\ga$, and $|0\ran$ is the photon vacuum state. The amplitude $d_\ga$ obeys the normalization condition
\begin{equation}\label{NormCond}
	\sum_{\ga} \bar d_\ga d_\ga =\sum_{\la}\int d\spk |d_\la(\spk)|^2=N_\ga,\qquad d_\la(\spk):=\sqrt{\frac{V}{(2\pi)^3}}d_\ga,
\end{equation}
where $N_\ga$ is the average number of photons in the coherent state, and is related to the average of a free electromagnetic field operator as
\begin{equation}
    \lan A_\mu(x)\ran=\sum_{\la=1,2} \int\frac{d\spk}{\sqrt{2(2\pi)^3k_0}} \big(d_\la(\spk) e_\mu^{(\la)}(\spk) e^{-ik_\nu x^
    \nu}+c.c. \big).
\end{equation}
The vectors $e_\mu^{(\la)}(\spk)$ specify the polarizations of physical photons. The density matrix of hadrons, $\hat R_h$, is arbitrary. Its representation in terms of creation-annihilation operators and its properties can be found, for example, in Appendix C of \cite{radet}.

Assume that only the state of a single photon is recorded at the final instant of time, $t_{out}\rightarrow\infty$. Such a measurement is described by the projector
\begin{equation}
	\hat \Pi_D=\hat \Pi_{ph}\otimes \hat 1_h\otimes\hat 1_{h^c},
\end{equation}
where $\hat \Pi_{ph}= 1-:e^{-\hat c^\dag D \hat c}:$ is the projector in the Fock space to the states containing at least one photon in the state specified by the projector $D_{\ga\bar\ga}$. The colons mean, as usual, the normal ordering. The projector $D$ in the one-particle Hilbert space is self-conjugate, $D^\dag=D$, and idempotent, $D^2=D$. We will also need the projector $\widetilde D:=1-D$.

Let us write all the leading contributions to the $S$-matrix that are admitted by the energy-momentum conservation law (see for details \cite{KazSol2022}),
\begin{equation}\label{SMatrix}
	\hat S=\hat 1+\hat W+\hat C+\ldots,
\end{equation}
where the ellipsis denotes the terms of higher order of smallness with respect to the coupling constant. The operator $\hat W$ is of second order in the coupling constant and describes Compton scattering by a hadron,
\begin{equation}
	\hat W=W^{\bar\ga\ga}_{\bar \al\al}\hat a^\dag_{\bar\al} \hat a_\al \hat c^\dag_{\bar\ga} \hat c_\ga.
\end{equation}
The explicit expression for the amplitude $W^{\bar\ga\ga}_{\bar\al\al}$ is given below in formula $\eqref{Ampl1}$. The operator $\hat C$ is also of the second order in the coupling constant and describes scattering of hadrons in the beam,
\begin{equation}
	\hat C=C_{\bar\al\bar\be\be\al}\hat a^\dag_{\bar\al}\hat a^\dag_{\bar\be}\hat a_{\be}\hat a_{\al}.
\end{equation}
The unitarity of $S$-matrix imposes the constraints on the operators $\hat W$ and $\hat C$ and the corresponding matrices $W^{\bar\ga\ga}_{\bar\al\al}$ and $C_{\bar\al\bar\be\be\al}$. Indeed,
\begin{equation}
	\hat S \hat S^\dag=(\hat 1+\hat W+\hat C+\ldots)(\hat 1+\hat W^\dag+\hat C^\dag+\ldots)=\hat 1+\hat W+\hat W^\dag+\hat C
    +\hat C^\dag+\ldots=\hat 1.
\end{equation}
Collecting the terms at the same monomials of creation-annihilation operators, we deduce that the operators $\hat W$ and $\hat C$ are anti-Hermitian in the leading order in the coupling constant,
\begin{equation}\label{AntiHermitWC}
	\hat W=-\hat W^\dag,\qquad \hat C=-\hat C^\dag,
\end{equation}
and their matrices satisfy the relations
\begin{equation}\label{AntiHermitWCMatr}
	W^{\bar\ga\ga}_{\bar\al\al}=-\bar{W}^{\ga\bar\ga}_{\al\bar\al},\qquad C_{\bar\al\bar\be\be\al}=-\bar{C}_{\al\be\bar\be\bar\al}.
\end{equation}

The inclusive probability to record a photon in the state singled out by the projector $D$ is found as \cite{KazSol2022,radet}
\begin{equation}\label{P_D_ini}
	P_D=\Sp(\hat \Pi_D \hat S \hat R \hat S^\dag).
\end{equation}
Then we substitute expansion \eqref{SMatrix} into this expression. Consider the term of the zeroth order in coupling constant
\begin{equation}
	\Sp(\hat \Pi_D \hat R)=\Sp(\hat \Pi_{ph}\hat R_{ph})=1-e^{-\bar d D d}\approx \bar d D d.
\end{equation}
The last approximate equality is obtained by developing the expression as a Taylor series with respect to $D$ and keeping only the leading contribution. This approximation is justified provided the projector $D$ projects onto the phase space domain of a small volume. For example, if $D$ is the projector to the state with definite momentum (see formula \eqref{DMomentum}), then it contains the infinitesimal phase space volume $d\spk'$. In what follows, we assume that $D$ singles out a phase space domain with a small volume.

Now we turn to the next contribution of the second order in the coupling constant stemming from the interaction of hadrons,
\begin{equation}
	\Sp(\hat\Pi_D\hat C\hat R)+\Sp(\hat\Pi_D\hat R \hat C^\dag)=\Sp(\hat \Pi_{ph}\hat R_{ph}) \big[\Sp(\hat C\hat R_h)
    +\Sp(\hat R_h\hat C^\dag)\big]=0.
\end{equation}
The last equality is valid in virtue of anti-Hermiticity condition \eqref{AntiHermitWC} and cyclic property of the trace. In other words, in the given order of perturbation theory, scattering of hadrons in the beam does not influence the inclusive probability to record a scattered photon.

Let us consider the Compton contribution in more detail. This contribution is written as
\begin{equation}
	\Sp(\hat\Pi_D\hat W\hat R)+\Sp(\hat\Pi_D\hat R \hat W^\dag)
    =\Sp(\hat \Pi_{ph}\hat c^\dag_{\bar \ga}\hat c_{\ga}\hat R_{ph})W_{\bar\al\al}^{\bar \ga \ga}\Sp(\hat a^\dag_{\bar \al}\hat a_{\al}\hat R_{h})+c.c.,
\end{equation}
where
\begin{equation}\label{SpurApp1}
	\Sp(\hat \Pi_{ph}\hat c^\dag_{\bar \ga}\hat c_{\ga}\hat R_{ph})=\big(1-e^{-\bar d D d}\big)
    (\bar d \widetilde D)_{\bar \ga}d_\ga+(\bar d D)_{\bar \ga}d_\ga\approx \bar d_{\bar \ga}d_\ga(\bar d D d)+(\bar d D)_{\bar \ga}d_\ga,
\end{equation}
and the approximate equality means that only the leading contribution in powers of the projector $D$ is retained. The trace,
\begin{equation}\label{density_oper}
    \Sp(\hat a^\dag_{\bar \al}\hat a_{\al}\hat R_{h})=\sum_{N=1}^{\infty}N\rho_{\al\bar\al}^{(N,1)} = \rho^{(1)}_{\al\bar{\al}},
\end{equation}
is the one-particle density matrix of hadrons (for its definition and a detailed exposition of its properties see, e.g., Appendix C of \cite{radet}). As a result, the Compton contribution becomes
\begin{equation}\label{PWR}
    \Sp(\hat\Pi_D\hat W\hat R)+\Sp(\hat\Pi_D\hat R \hat W^\dag)=(\bar d D d)W^{\bar\ga\ga}_{\bar\al\al}\bar d_{\bar\ga}d_\ga \rho^{(1)}_{\al\bar{\al}} +W^{\bar\ga\ga}_{\bar\al\al}(\bar d D)_{\bar\ga}d_\ga \rho^{(1)}_{\al\bar{\al}}+c.c.
\end{equation}
It follows from the anti-Hermiticity of the operator $\hat{W}$ and Hermiticity of the operator $D$ and the one-particle density matrix $\rho_{\al\bar\al}^{(1)}$ that the contribution of the first term on the right-hand side of \eqref{PWR} is equal to zero on accounting the complex conjugate terms. Then the inclusive probability to record a single photon reads
\begin{equation}\label{P_D}
    P_D=D_{\bar\ga\ga} [\rho^{ph}_{\ga\bar\ga} +(\Phi_{\ga \ga'}\rho^{ph}_{\ga'\bar\ga} +h.c.)],
\end{equation}
where we have introduced the one-particle density matrix of photons $\rho^{ph}_{\ga\bar\ga}=d_\ga \bar d_{\bar\ga}$ and the amplitude of coherent Compton scattering
\begin{equation}\label{amplitude_coh_Compt}
    \Phi_{\ga\ga'}:=W^{\ga \ga'}_{\bar\al\al}\rho_{\al\bar\al}^{(1)}.
\end{equation}
The inclusive probability \eqref{P_D} gives the hologram of the one-particle density matrix of hadrons. The factor in the square brackets in \eqref{P_D},
\begin{equation}\label{photon_dens_mat_out}
    (\rho^{ph}_{out})_{\ga\bar\ga}= \rho^{ph}_{\ga\bar\ga}+ (\Phi_{\ga \ga'}\rho^{ph}_{\ga'\bar\ga} +h.c.),
\end{equation}
can be regarded as the one-particle density matrix of photons scattered by hadrons, the state of hadrons being not detected. It is not difficult to see from \eqref{photon_dens_mat_out} that if the initial one-particle density matrix of photons described a pure state, then this density matrix also describes a pure state in the given order of perturbation theory.

Formula \eqref{photon_dens_mat_out} is also valid for a one-particle initial state of photons. This can be verified by conducting the above calculations once again for the one-particle initial state of photons or by taking the leading (quadratic) contribution to \eqref{P_D_ini} as $d_\ga\rightarrow0$ (see the expression for the coherent state \eqref{R_ph} and the traces with the photon density matrix \eqref{SpurApp1}). The density matrix $\rho^{ph}_{\ga'\bar\ga}$ can describe a mixed state for a one-particle state of photons. Below we will also consider this general case.

The hologram of the one-particle density matrix of hadrons can be written in terms of the one-particle density matrix of scattered photons as
\begin{equation}
    P_D=D_{\bar\ga\ga} (\rho^{ph}_{out})_{\ga\bar\ga}.
\end{equation}
Let $D$ be the projector to the state of a photon with definite momentum $\spk'$ and polarization characterized by the Stokes vector $\boldsymbol\zeta'$,
\begin{equation}\label{DMomentum}
	D_{\bar\ga\ga}=\frac{(2\pi)^3}{V}D^{(\boldsymbol\zeta')}_{\la_2\la_1}\delta(\spk_2-\spk')\delta(\spk_1-\spk')d\spk',
\end{equation}
where
\begin{equation}
    D^{(\boldsymbol\zeta')}_{\la_2\la_1}=\frac{1}{2}[1+(\boldsymbol\sigma\boldsymbol\zeta')]_{\la_2\la_1},
\end{equation}
and $\s^i$ are the Pauli matrices. Taking into account the normalization of states \eqref{NormCond}, the inclusive probability can be cast into the form
\begin{equation}\label{inclus_prob}
	dP(\boldsymbol\zeta',\spk') =\sum_{\la_1,\la_2}D_{\la_2\la_1}^{(\boldsymbol\zeta')}(\rho^{ph}_{out})_{\la_1\la_2}(\spk',\spk')d\spk'.
\end{equation}
Decomposing the one-particle density matrix of scattered photons with respect to the basis of $\s$-matrices,
\begin{equation}\label{D_rho_Pauli}
	(\rho^{ph}_{out})_{\la_1\la_2}(\spk',\spk')=\frac{\rho^{ph}_{out}(\spk',\spk')}{2}
    [1+(\boldsymbol\sigma\boldsymbol\xi^{ph}_{out})]_{\la_1\la_2},
\end{equation}
where $\boldsymbol\xi^{ph}_{out}=\boldsymbol\xi^{ph}_{out}(\spk',\spk')$ is the Stokes vector of scattered photons, we arrive at
\begin{equation}
	dP(\boldsymbol\zeta',\spk')=\frac{\rho^{ph}_{out}}{2}(1+(\boldsymbol\zeta'\boldsymbol\xi^{ph}_{out}))d\spk'.
\end{equation}
Further, we will obtain the explicit expressions for $\rho^{ph}_{out}$ and $\bs\xi^{ph}_{out}$.

\section{Density matrix of scattered photons}\label{Density_Matrix_of_Scattered_Photons}

The amplitude of the Compton process takes the form \cite{PeskSchr,KazSol2022}
\begin{equation}\label{Ampl1}
	W_{\bar\al\al}^{\bar \ga \ga}=-i (2\pi)^4 \delta(p'+k'-p-k)
    \frac{M \bar{e}^{(\la')}_\mu(\spk') e_{\nu}^{(\la)}(\spk)}{2V^2\sqrt{k_0'k_0p_0'p_0}}\bar u^{s'}(\spp')\Ga^{\mu\nu}u^s(\spp),
\end{equation}
where
\begin{equation}\label{Gamma_munu}
	\bar u^{s'}(\spp')\Ga^{\mu\nu}u^s(\spp):=\bar u^{s'}(\spp')\left[\Ga^\mu(-k')\frac{\hat p_c+
    \hat k_c+M}{(p_c+k_c)^2-M^2}\Ga^\nu(k)+\Ga^\nu(k)\frac{\hat p_c-\hat k_c+M}{(p_c-k_c)^2-M^2}\Ga^\mu(-k')\right]u^s(\spp),
\end{equation}
and
\begin{equation}\label{pc_kc_q}
    p^\mu=p_c^\mu+q^\mu/2,\qquad p'^\mu=p_c^\mu-q^\mu/2,\qquad k^\mu=k_c^\mu-q^\mu/2,\qquad k'^\mu=k_c^\mu+q^\mu/2.
\end{equation}
The sign at the momentum in the argument of the vertex $\Ga^\mu$ indicates that the photon momentum inflows into the vertex ($+$) or outflows from it ($-$). The representation of the amplitude in term of the momenta $p_c$, $k_c$ and the transferred momentum $q$ is useful in the small recoil limit where $q$ is small. There are the relations
\begin{equation}
	(p_cq)=0,\quad (k_cq)=0,\quad p_c^2=M^2-q^2/4,\quad k_c^2=-q^2/4.
\end{equation}

Taking into account the explicit form of the Compton scattering amplitude \eqref{Ampl1}, we write the one-particle density matrix of scattered photons \eqref{photon_dens_mat_out} entering into the hologram \eqref{inclus_prob} as
\begin{equation}\label{photon_dens_mat_out_expan}
    (\rho^{ph}_{out})_{\la_1\la_2}(\spk',\spk') = \overset{(0)}{\rho}{}^{ph}_{\la_1\la_2} +\overset{(2)}{\rho}{}^{ph}_{\la_1\la_2} ,\qquad \overset{(2)}{\rho}{}^{ph}_{\la_1\la_2}(\spk',\spk') =r_{\la_1\la_2}+ \bar{r}_{\la_2\la_1},
\end{equation}
where
\begin{equation}\label{rho_out}
\begin{split}
	\overset{(0)}{\rho}{}^{ph}_{\la_1\la_2}&=\rho^{ph}_{\la_1\la_2}(\spk',\spk'),\\
	r_{\la_1\la_2}&=-\frac{iM}{8\pi^2}\sum_{\la,s,s'}\int d\spk d\spp d\spp'
    \rho^{ph}_{\la\la_2}(\spk,\spk') \rho^{(1)}_{ss'}(\spp,\spp') \delta(p'+k'-p-k) \frac{\bar e^{(\la_1)}_\mu(\spk') e_\nu^{(\la)}(\spk)\bar u^{s'}(\spp')\Ga^{\mu\nu}u^s(\spp)}{\sqrt{k'_0k_0p'_0p_0}}.
\end{split}
\end{equation}
Hereinafter, we use the following normalization of one-particle density matrices
\begin{equation}\label{normalization_dens_matr}
    \sum_\la\int d\spk \rho^{ph}_{\la\la}(\spk,\spk)=N_\ga,\qquad \sum_s \int d\spp \rho^{(1)}_{ss}(\spp,\spp)=N_h,
\end{equation}
where $N_\ga$ and $N_h$ are the average numbers of photons and hadrons in the initial state, respectively. Substituting the expression for the electromagnetic vertex \eqref{hadron_curr} into the square brackets in \eqref{Gamma_munu}, we have
\begin{equation}
	\Ga^{\mu\nu}=e^2_p\Ga_1^{\mu\nu}+e_p\mu_p\Ga_2^{\mu\nu}+\mu_p^2\Ga_3^{\mu\nu},
\end{equation}
where
\begin{equation}
\begin{split}
    \Ga^{\mu\nu}_1:=&\,\ga^\mu\frac{\hat p_c+\hat k_c+M}{(p_c+k_c)^2-M^2}\ga^\nu+\ga^\nu\frac{\hat p_c-\hat k_c+M}{(p_c-k_c)^2-M^2}\ga^\mu,\\
    \Ga^{\mu\nu}_2:=&\,\ga^\mu\frac{\hat p_c+\hat k_c+M}{(p_c+k_c)^2-M^2}i\sigma^{\nu\rho}k_\rho-i\sigma^{\mu\la}k'_\la\frac{\hat p_c+\hat k_c+M}{(p_c+k_c)^2-M^2}\ga^\nu-\\
    &-\ga^\nu\frac{\hat p_c-\hat k_c+M}{(p_c-k_c)^2-M^2}i\sigma^{\mu\la}k'_\la+i\sigma^{\nu\rho}k_\rho\frac{\hat p_c-\hat k_c+M}{(p_c-k_c)^2-M^2}\ga^\mu,\\
    \Ga^{\mu\nu}_3:=&\,\sigma^{\mu\la}k'_\la\frac{\hat p_c+\hat k_c+M}{(p_c+k_c)^2-M^2}\sigma^{\nu\rho}k_\rho+\sigma^{\nu\rho}k_\rho\frac{\hat p_c-\hat k_c+M}{(p_c-k_c)^2-M^2}\sigma^{\mu\la}k'_\la.
\end{split}
\end{equation}
In order to evaluate the matrices $\Ga^{\mu\nu}_n$ standing between the Dirac spinors, we employ formula (32) of \cite{KRS2023},
\begin{equation}
	\bar u^{s'}(\spp')\Ga^{\mu\nu} u^s(\spp)=\frac{\Sp\Big[\Ga^{\mu\nu} (M+\hat p)
		[\de_{s's} +(\s_a)_{s's}\tau_{ai}\ga^i\ga^5]\frac{1+\ga^0}{2}
		(M+\hat p')\Big]}{4M\sqrt{(p_0+M)(p'_0+M)}},
\end{equation}
where $\s_a$ are the Pauli matrices and
\begin{equation}
\begin{gathered}
    \bs{\tau}_1=\re\mathbf{f},\qquad \bs{\tau}_2=\im\mathbf{f},\qquad\bs{\tau}_3=\bs\tau,\\
    \mathbf{f}=e^{i\vf}(\cos\theta \cos\vf-i\sin\vf,\cos\theta \sin\vf+i\cos\vf,-\sin\theta).
\end{gathered}
\end{equation}
Let us introduce a convenient notation
\begin{equation}\label{spin_curr_rep}
	\bar u^{s'}(\spp')\Ga_n^{\mu\nu}u^s(\spp)=\delta_{s's}G_n^{\mu\nu}(\spp,\spp')-(\s_a)_{s's}\tau_{ai}Z_n^{i\mu\nu}(\spp,\spp'), \quad
    n=\overline{1,3},
\end{equation}
where
\begin{equation}\label{G_n_Z_n_defn}
\begin{split}
	G_n^{\mu\nu}&:=\frac{1}{4M\sqrt{(p_0+M)(p'_0+M)}}\Sp{\left[\Ga^{\mu\nu}_n(M+\hat p)\frac{1+\ga_0}{2}(M+\hat p')\right]},\\
	Z_n^{i\mu\nu}&:=\frac{-1}{4M\sqrt{(p_0+M)(p'_0+M)}}\Sp{\left[\Ga^{\mu\nu}_n(M+\hat p)\ga^{i}\ga^5\frac{1+\ga_0}{2}(M+\hat p')\right]}.
\end{split}
\end{equation}
The explicit expressions for the tensors $G_1^{\mu\nu}$ and $Z_1^{i\mu\nu}$ taken out of the mass-shell were obtained in \cite{AKS2025}.

It is useful to represent the product of the polarization vectors in the form analogous to \eqref{spin_curr_rep},
\begin{equation}
	e^{(\la)}_\mu(\spk)\bar{e}^{(\la')}_\nu(\spk')=\frac12[\delta_{\la'\la}M_{\mu\nu}(k,k')-(\s_a)_{\la'\la}N_{a\mu\nu}(k,k')],
\end{equation}
where
\begin{equation}\label{polar_vec_prods}
	M_{\mu\nu}(k,k'):=\sum_\la e^{(\la)}_\mu(\spk)\bar{e}^{(\la)}_\nu(\spk'),
    \qquad N_{a\mu\nu}(k,k'):=-\sum_{\la,\la'} (\s_a)_{\la\la'} e^{(\la)}_\mu(\spk)\bar{e}^{(\la')}_\nu(\spk').
\end{equation}
Some properties of the tensors $M_{\mu\nu}$ and $N_{a\mu\nu}$ are collected in Appendix \ref{M_and_N_App}.

As the physical polarization vectors, we take the linear polarization vectors that are constructed in the following way. Let us choose some timelike $4$-vector $t^\mu$ distinguishing the laboratory reference frame. Notice that this vector cannot be the $4$-momentum of a hadron, $p^\mu$, as long as the integration over $p^\mu$ is performed in the wave packet. Let us also choose some spacelike $4$-vector $d^\mu$, $(td)=0$, $d^2=-1$, defining the direction in space with respect to which the polarization vectors will be specified. Construct the $4$-vector orthogonal to $t^\mu$ and $d^\mu$ from the $4$-momentum of a photon as
\begin{equation}
    k_\perp^\mu:=k^\mu-(kt)t^\mu +(kd)d^\mu, \qquad k_\perp:=\sqrt{-k_\perp^2}=\sqrt{(kt)^2-(kd)^2}.
\end{equation}
Then the polarization vectors can be cast into the form
\begin{equation}\label{polarization_vects}
	e^{(1)}_{\mu}(\spk)=\frac{t_{\mu}(kd)-d_\mu(kt)-k_\mu (kd)/(kt)}{k_\perp},\qquad
    e^{(2)}_{\mu}(\spk)=\frac{\e_{\mu\nu\rho\la}t^\nu k^\rho d^{\la}}{k_\perp}.
\end{equation}
These vectors together with the vector $k^\mu/(kt)-t^\mu$ constitute a right-handed orthonormal triple and possess the properties
\begin{equation}
	k^\mu e^{(\la)}_\mu(\spk)=0,\qquad (e^{(\la)}(\spk)\bar e^{(\la')}(\spk))=-\delta_{\la\la'},\qquad t^\mu e^{(\la)}_\mu(\spk)=0.
\end{equation}
The explicit expressions for the tensors \eqref{polar_vec_prods} constructed by using the polarization vectors \eqref{polarization_vects} are presented in Appendix \ref{M_and_N_App}.

In the present paper, in order to simplify the calculations, we restrict ourselves to the case of elastic scattering of photons by electrically neutral hadrons (neutrons) with anomalous magnetic moment $\mu_p$. Decomposing the one-particle density matrices of neutrons and photons in the basis of $\s$-matrices,
\begin{equation}\label{dens_mat_ini}
    \rho^{(1)}_{ss'}(\spp,\spp')=\frac{\rho^{(1)}(\spp,\spp')}2 [1+(\bs\sigma\bs\xi^h(\spp,\spp'))]_{ss'},\qquad \rho^{ph}_{\la\la_2}(\spk,\spk')=
    \frac{\rho^{ph}(\spk,\spk')}{2}[1+(\bs\s\bs\xi^{ph}(\spk,\spk'))]_{\la\la_2},
\end{equation}
we deduce that the second term in the expansion of the one-particle spin density matrix of scattered photons \eqref{photon_dens_mat_out_expan} with respect to the coupling constant is reduced to
\begin{equation}\label{rho2_ph}
\begin{split}
	\overset{(2)}{\rho}{}^{ph'}_{\la_1\la_2}=&\,\frac{M\mu_p^2}{16\pi^2}\int
    \frac{d\spk d\spp d\spp'}{\sqrt{k'_0k_0p'_0p_0}}
	\delta(p'+k'-p-k)\Big\{\delta_{\la_1\la_2}\im\big[\rho^{ph}\rho^{(1)}
    (M_{\nu\mu}-\xi^{ph}_b N_{b\nu\mu})(G^{\mu\nu}_3-\xi^h_iZ^{i\mu\nu}_3
    )\big]+\\
    &+(\s_a)_{\la_1\la_2}\im\big[\rho^{ph}\rho^{(1)}(\xi^{ph}_a M_{\nu\mu}-N_{a\nu\mu}+i\e_{abc}\xi^{ph}_b N_{c\nu\mu})(G^{\mu\nu}_3-\xi^h_iZ^{i\mu\nu}_3 )\big]\Big\},
\end{split}
\end{equation}
where $\xi^h_i(\spp,\spp'):=\xi^h_a(\spp,\spp')\tau_{ai}$ and all the omitted arguments of the objects entering into this formula are the same as when they were introduced for the first time.

The interference term \eqref{rho2_ph} is substantially different from zero only for small transferred momenta, the transferred momentum being of order of the standard deviation of momenta in the one-particle density matrix of photons $|\spq|\lesssim\sigma_{ph}$. Further, we will perform all the calculation in the small recoil limit
\begin{equation}
    |\spk'|\gg|\spq|,\qquad |\spp'|\gg|\spq|,
\end{equation}
where, in the latter estimate, $|\spp'|$ denotes a typical value of modulus of momenta in the one-particle density matrix of hadrons. The integral over the momentum $\spp$ in \eqref{rho2_ph} is removed by the three-dimensional delta function and
\begin{equation}
    \spp=\spp'+\spk'-\spk=\spp'+\spq.
\end{equation}
Since in the small recoil limit
\begin{equation}
    k_0'-k_0\approx(\spn'\spq),\qquad p_0'-p_0\approx -(\bs\be'_h\spq),
\end{equation}
where $\spn'=\spk'/k_0'$ and $\bs\be'_h=\spp'/p'_0$, the delta function expressing the energy conservation law, can be represented in the form
\begin{equation}
	\delta(p_0'+k_0'-p_0-k_0)\approx \de(\Delta \bs{\be}\spq),
\end{equation}
where $\Delta\bs\be:=\spn'-\bs\be'_h$. Now we introduce the components of momenta longitudinal and transverse to $\Delta \bs\be$. For example,
\begin{equation}
	\spq=\spq_\perp+q_\parallel \frac{\De \bs\be}{|\De \bs\be|},\quad q_\parallel=\frac{(\De\bs\be \spq)}{|\De\bs\be|}.
\end{equation}
As a result,
\begin{equation}\label{de-func_energy}
	\de(p_0'+k_0'-p_0-k_0)\approx\de(q_\parallel)/|\De\bs\be|.
\end{equation}
Change the integration variable in \eqref{rho2_ph} from $\spk$ to $\spq$. Then the delta function \eqref{de-func_energy} removes integration with respect to the longitudinal component of $\spq$ and sets $q_\parallel=0$.

Suppose for simplicity that the neutrons are nonrelativistic, viz., $|\bs\be_h'|\ll1$. In that case, $|\De\bs\be|\approx|\spn'|=1$ and
\begin{equation}\label{neutron_dens_mat2}
\begin{split}
	\overset{(2)}{\rho}{}^{ph'}_{\la_1\la_2}=&\,\frac{\mu_p^2}{16\pi^2 k_0'}\int d\spq_\perp d\spp'
	\Big\{\delta_{\la_1\la_2}\im\big[\rho^{ph}\rho^{(1)} (M_{\nu\mu}-\xi^{ph}_b N_{b\nu\mu})(G^{\mu\nu}_3
    -\xi^h_iZ^{i\mu\nu}_3 )\big]+\\
    &+(\s_a)_{\la_1\la_2}\im\big[\rho^{ph}\rho^{(1)}(\xi^{ph}_a M_{\nu\mu}-N_{a\nu\mu}+i\e_{abc}\xi^{ph}_b N_{c\nu\mu})(G^{\mu\nu}_3-\xi^h_iZ^{i\mu\nu}_3 )\big]\Big\}\Big|_{\substack{\spp=\spp'+\spq_\perp\\\spk=\spk'-\spq_\perp}},
\end{split}
\end{equation}
where only the leading with respect to $q^\mu$ contributions to the contractions of $M_{\nu\mu}$, $N_{a\nu\mu}$ with $G_3^{\mu\nu}$ and $Z_3^{i\mu\nu}$ are retained (see Appendix \ref{Contractions_App}). The dependence of the density matrices on $\spq_\perp$ is taken into account exactly. The interference term \eqref{neutron_dens_mat2} is written as
\begin{equation}
	\overset{(2)}{\rho}{}^{ph'}_{\la_1\la_2}=\frac12[a\delta_{\la_1\la_2}+b_a (\s_a)_{\la_1\la_2}],
\end{equation}
where the expression for $a$ and $b_a$ follow immediately from \eqref{neutron_dens_mat2}. Then the one-particle spin density matrix of scattered photons can be cast into the form \eqref{D_rho_Pauli}, where
\begin{equation}\label{rho_ph_out}
	\rho^{ph}_{out}=\rho^{ph}+a,
    \qquad \bs\xi^{ph}_{out}=\frac{\rho^{ph}\bs\xi^{ph}+\spb}{\rho^{ph}+a}\approx (1-a/\rho^{ph})\bs\xi^{ph}+\spb/\rho^{ph},
\end{equation}
in the leading order of perturbation theory and $\rho^{ph}\equiv\rho^{ph}(\spk',\spk')$, $\bs\xi^{ph}\equiv\bs\xi^{ph}(\spk',\spk')$. Notice that if $\bs\xi^{ph}=const$ and $|\bs\xi^{ph}|=1$, then \eqref{rho2_ph} implies that
\begin{equation}\label{bxi_a}
    (\bs\xi^{ph}\spb)=a.
\end{equation}
This property is a particular case of the general fact mentioned after formula \eqref{photon_dens_mat_out} stating that if the initial one-particle density matrix described a pure state, then the one-particle density matrix of scattered photons also describes a pure state in the leading nontrivial order of perturbation theory. Indeed, it follows from \eqref{rho_ph_out} on accounting \eqref{bxi_a} and $|\bs\xi^{ph}|=1$ that
\begin{equation}
    (\bs\xi^{ph}_{out})^2\approx1-2(a- (\bs\xi^{ph}\spb))/\rho^{ph}=1,
\end{equation}
i.e., the state of a photon after scattering is pure with respect to spin.

\section{Evolution of the Stokes parameters}\label{Evolution_of_Stokes_Parameters}

Let us consider expression \eqref{neutron_dens_mat2} in more detail for the particular cases of polarized and unpolarized neutrons and photons. For unpolarized incoming particles, $\bs\xi^{ph}=\bs\xi^h=0$, we have
\begin{equation}
	\overset{(2)}{\rho}{}^{ph'}_{\la_1\la_2}=\frac{\mu_p^2}{16\pi^2 k_0'}\int d\spq_\perp d\spp'
	\Big\{\delta_{\la_1\la_2}\im\big[\rho^{ph}\rho^{(1)} M_{\nu\mu} G^{\mu\nu}_3
    -(\s_a)_{\la_1\la_2} \im\big[\rho^{ph}\rho^{(1)} N_{a\nu\mu} G^{\mu\nu}_3\big]
    \Big\}\Big|_{\substack{\spp=\spp'+\spq_\perp\\\spk=\spk'-\spq_\perp}}.
\end{equation}
The contractions arising in the integrand are given in Appendix \ref{Contractions_App}. Substituting them, we arrive at
\begin{equation}\label{non_pol_dens_out}
\begin{split}
	a&=-\frac{\mu_p^2}{2M k'_0}\int \frac{d\spq_\perp d\spp'}{(2\pi)^2}\spq_\perp^2\im[\rho^{ph}\rho^{(1)}],\\
	b_1&=\frac{\mu_p^2}{M k'_0}\frac{d_l\e_{ijk}d_i n'_j}{n'^2_{\perp}}\int
    \frac{d\spq_\perp d\spp'}{(2\pi)^2}q^\perp_l q^\perp_k \im[\rho^{ph}\rho^{(1)}],\\
	b_2&=-\frac{\mu_p^2}{2M}\e_{ijk}n'_i \int
    \frac{d\spq_\perp d\spp'}{(2\pi)^2} \be'^h_j q^\perp_k \re[\rho^{ph}\rho^{(1)}],\\
	b_3&=a+\frac{\mu_p^2}{M k'_0}\frac{d_i d_j}{n'^2_{\perp}}\int
    \frac{d\spq_\perp d\spp'}{(2\pi)^2} q^\perp_i q^\perp_j \im[\rho^{ph}\rho^{(1)}],
\end{split}
\end{equation}
in the leading order of the small recoil approximation, where $n^\mu:=k^\mu/k_0$ and $n_\perp:=k_\perp/k_0$.

For polarized incoming particles, when $|\bs\xi^{ph}|\gg|\spq|/M$, $|\bs\xi^h|\gg|\spq|/M$, the contractions of tensors in the integrand of \eqref{neutron_dens_mat2} are also presented in Appendix \ref{Contractions_App}. The main contribution in the small recoil limit comes from the contraction with the tensor $N_{2\mu\nu}$. In this case, we have
\begin{equation}\label{pol_dens_out}
\begin{split}
	a&=\mu_p^2\int \frac{d\spq_\perp d\spp'}{2\pi^2}\im[\rho^{ph}\rho^{(1)}\xi_2^{ph}(n's_h)],\\
	b_1&=\mu_p^2\int \frac{d\spq_\perp d\spp'}{2\pi^2}\re[\rho^{ph}\rho^{(1)}\xi_3^{ph}(n's_h)],\\
	b_2&=\mu_p^2\int \frac{d\spq_\perp d\spp'}{2\pi^2}\im[\rho^{ph}\rho^{(1)}(n's_h)],\\
	b_3&=-\mu_p^2\int \frac{d\spq_\perp d\spp'}{2\pi^2}\re[\rho^{ph}\rho^{(1)}\xi_1^{ph}(n's_h)],
\end{split}
\end{equation}
where recall that
\begin{equation}
\begin{aligned}
	\rho^{ph}&=\rho^{ph}(\spk'-\spq_\perp,\spk'),&\qquad \bs\xi^{ph}&=\bs\xi^{ph}(\spk'-\spq_\perp,\spk'),\\
    \rho^{(1)}&=\rho^{(1)}(\spp'+\spq_\perp,\spp'),&\qquad \bs\xi^h&=\bs\xi^h(\spp'+\spq_\perp,\spp),
\end{aligned}
\end{equation}
and the $4$-vector of a neutron spin has been introduced as
\begin{equation}\label{s_h_mu_defn}
    s^\mu_h:=\Big(\frac{(\bs\xi^h\spp')}{M},\bs\xi^h+\frac{\spp'(\bs\xi^h\spp')}{M(p'_0+M)}\Big).
\end{equation}
In the non-relativistic limit, $s_h^\mu=(0,\bs\xi^h)$.

In the case $\bs\xi^{ph}=const$, formulas \eqref{rho_ph_out} and \eqref{pol_dens_out} imply the expressions for a change of the photon Stokes vector on scattering by polarized neutrons
\begin{equation}\label{de_xi}
    \de\xi^{ph}_2=\vk''\big(1-(\xi^{ph}_2)^2\big),\qquad \de\xi^{ph}_+=i(\vk'+i\vk''\xi^{ph}_2)\xi^{ph}_+,
\end{equation}
where $\xi^{ph}_\pm=\xi^{ph}_3\pm i\xi^{ph}_1$ and
\begin{equation}
    \vk':=\frac{\mu_p^2}{\rho^{ph}(\spk',\spk')}\int \frac{d\spq_\perp d\spp'}{2\pi^2}\re[\rho^{ph}\rho^{(1)}(n's_h)],\qquad
    \vk'':=\frac{\mu_p^2}{\rho^{ph}(\spk',\spk')}\int \frac{d\spq_\perp d\spp'}{2\pi^2}\im[\rho^{ph}\rho^{(1)}(n's_h)].
\end{equation}
Equations \eqref{de_xi} describe the evolution of the photon Stokes vector in scattering by polarized neutrons. The parameter $\vk''$ determines a change in the degree of photon circular polarization while $\vk'$ specifies the magnitude of rotation of its linear polarization plane (for the interpretation of the Stokes parameters see, e.g., \cite{BaKaStrbook}). It is seen that the photon acquires a circular polarization in multiple scattering on identically polarized neutrons, viz.,
\begin{equation}\label{xi_2_limit}
    \xi^{ph}_2\rightarrow\sgn(\vk'').
\end{equation}
The photon state becomes pure in the spin degrees of freedom. It is clear from equations \eqref{de_xi} that
\begin{equation}
    \de\big[1-(\xi^{ph}_2)^2-\xi^{ph}_+\xi^{ph}_-\big]^{1/2}=-\vk''\xi^{ph}_2\big[1-(\xi^{ph}_2)^2-\xi^{ph}_+\xi^{ph}_-\big]^{1/2}.
\end{equation}
In particular, this equation demonstrates once more that a pure photon state remains pure in coherent scattering in the leading nontrivial order of perturbation theory.

To conclude this section, we consider coherent elastic scattering of polarized photons by unpolarized neutrons, $\bs\xi^h=0$.  In the small recoil approximation, the contraction $N_{2\nu\mu}G_3^{\mu\nu}$ dominates among the other contractions in the integrand of \eqref{neutron_dens_mat2} (see Appendix \ref{Contractions_App}). Consequently, a change in the photon Stokes vector caused by scattering takes the form
\begin{equation}\label{half_pol_dens_out}
\begin{split}
	a&=-\frac{\mu_p^2}{2M}\e_{ijk}n'_i \int \frac{d\spq_\perp d\spp'}{(2\pi)^2}\be'^h_j
    q^\perp_k\re[\rho^{ph}\rho^{(1)}\xi_2^{ph}],\\
	b_1&=\frac{\mu_p^2}{2M}\e_{ijk}n'_i \int \frac{d\spq_\perp d\spp'}{(2\pi)^2}\be'^h_j
    q^\perp_k\im[\rho^{ph}\rho^{(1)}\xi_3^{ph}],\\
	b_2&=-\frac{\mu_p^2}{2M}\e_{ijk}n'_i \int \frac{d\spq_\perp d\spp'}{(2\pi)^2}\be'^h_j
    q^\perp_k\re[\rho^{ph}\rho^{(1)}],\\
	b_3&=-\frac{\mu_p^2}{2M}\e_{ijk}n'_i \int \frac{d\spq_\perp d\spp'}{(2\pi)^2}\be'^h_j
    q^\perp_k\im[\rho^{ph}\rho^{(1)}\xi_1^{ph}].
\end{split}
\end{equation}
For $\bs\xi^{ph}=const$, a change of the Stokes vector is described by \eqref{de_xi} with
\begin{equation}
    \vk'= \frac{\mu_p^2 \e_{ijk}n'_i}{2M\rho^{ph}(\spk',\spk')} \int \frac{d\spq_\perp d\spp'}{(2\pi)^2}\be'^h_j
    q^\perp_k\im[\rho^{ph}\rho^{(1)}],\qquad
    \vk''=-\frac{\mu_p^2 \e_{ijk}n'_i}{2M\rho^{ph}(\spk',\spk')} \int \frac{d\spq_\perp d\spp'}{(2\pi)^2}\be'^h_j
    q^\perp_k\re[\rho^{ph}\rho^{(1)}].
\end{equation}
All the above conclusions about the evolution of the Stokes vector are also valid in this case. In particular, the scattered photon possesses the circular polarization \eqref{xi_2_limit} after multiple coherent scattering by neutrons provided the neutrons are in the quantum states such that $\vk''$ is of the same sign for the majority of neutrons on which the photon is scattered. For $\vk'' = 0$, only the rotation of the linear polarization plane of scattered photons occurs and the degree of their circular polarization is conserved.

\section{Gaussian wave packets}\label{Gaussian_Wave_Packets}

In this section, we consider a special case of coherent elastic scattering of photons by neutrons where the initial states of the photons and neutrons have the one-particle density matrices \eqref{dens_mat_ini} of the Gaussian form
\begin{equation}\label{Gaussian_states}
\begin{split}
    \rho^{ph}(\spk,\spk')&=\frac{N_\ga}{(2\pi)^{3/2}\sigma^3_{ph}}\exp\Big[-\frac{(\spk-\spk_0)^2}{4\sigma^2_{ph}} -\frac{(\spk'-\spk_0)^2}{4\sigma^2_{ph}}-i\spg_{ph}(\spk-\spk')\Big],\\
    \rho^{(1)}(\spp,\spp')&=\frac{N_h}{(2\pi)^{3/2}\sigma^3_{h}}\exp\Big[-\frac{(\spp-\spp_0)^2}{4\sigma^2_{h}} -\frac{(\spp'-\spp_0)^2}{4\sigma^2_{h}}-i\spg_{h}(\spp-\spp')\Big],
\end{split}
\end{equation}
where $\s^2_{ph}$ and $\s^2_h$ are the momentum dispersions in the photon and neutron one-particle density matrices, $\spk_0$ and $\spp_0$ are the average momenta of the photon and neutron in the initial state, $\spg_{ph}$ and $\spg_h$ are the average positions of photons and neutrons in the coordinate space at the instant of time $t=0$. We will suppose that the Stokes vectors $\bs\xi^{ph}$ and $\bs\xi^h$ are independent of the momenta of the one-particle density matrices.

\begin{figure}[tp]
\centering
\includegraphics*[width=0.45\linewidth]{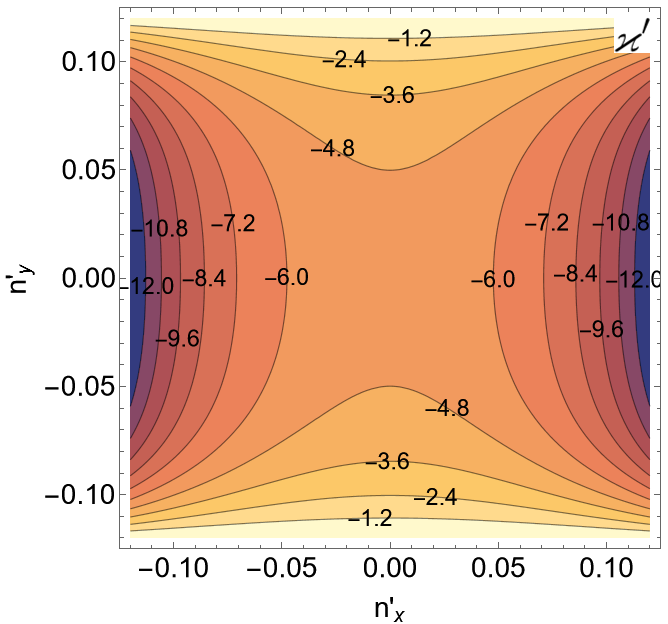}\;
\includegraphics*[width=0.45\linewidth]{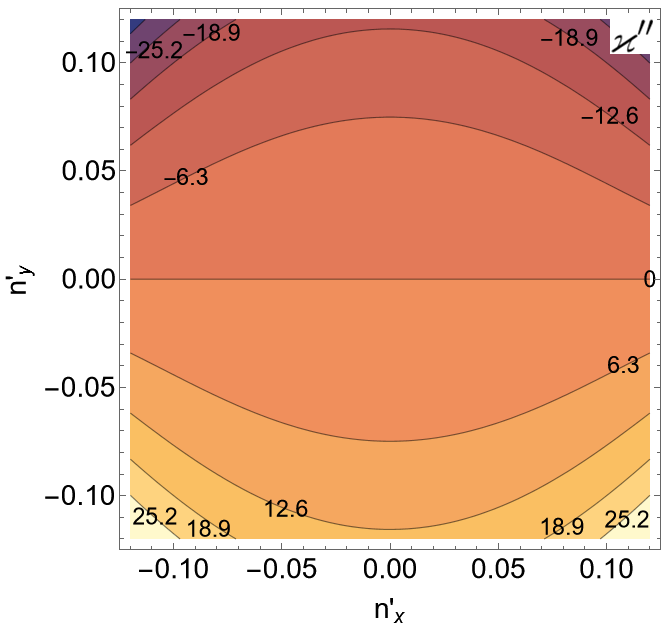}
\caption{{\footnotesize The parameters $\vk'$ and $\vk''$ multiplied by $10^{13}/N_h$ as functions of the direction of emission of the scattered photon $(n'_x,n'_y)$ for coherent elastic scattering of polarized photons by polarized neutrons with $\bs\xi_h=(0,0,1)$. The parameters $\vk'$ and $\vk''$ are defined in \eqref{de_xi} and determine the evolution of the Stokes vector of scattered photons. The states of photons and neutrons are taken in the form \eqref{Gaussian_states} with the parameters $\spk_0=(0,0,100)$ keV, $\s_{ph}=5$ keV, $\mathbf{g}_{ph}-\mathbf{g}_h=(0,0.2,0)$ keV${}^{-1}$, $\spp_0=0$ keV, $\s_h=5$ keV. }}
\label{kappa_polarized_plots}
\end{figure}

The Gaussian type integrals entering the one-particle density matrix of scattered photons with parameters \eqref{non_pol_dens_out}, \eqref{pol_dens_out}, \eqref{half_pol_dens_out} are easily performed by using the generating function
\begin{equation}
\begin{split}
    I(\mathbf{J}_\perp,\mathbf{K}):=&\int d\spq_\perp d\spp'\rho^{ph}(\spk'-\spq_\perp,\spk')\rho^{(1)}(\spp'+\spq_\perp,\spp') e^{i\spq_\perp\mathbf{J}_\perp +i\spp'\mathbf{K} }=\\
    =&\frac{2\Si^2N_{\ga}N_h}{\sqrt{2\pi} \s_{ph}^3} \exp\Big[-\frac{\De\spk^2}{2\s_{ph}^2} +\Si^2
    \Big(\mathbf{v}_\perp + i\mathbf{J}_\perp -
    \frac{i\mathbf{K}_\perp}{2}\Big)^2 +i\spp_0\mathbf{K} -\frac{\s_h^2}{2}\mathbf{K}^2 \Big],
\end{split}
\end{equation}
where $\Sigma^{-2}:=\sigma_{ph}^{-2}+\sigma_{h}^{-2}/2$, $\mathbf{v}_\perp:=i(\spg^{ph}_\perp-\spg^{h}_\perp)-\spk^0_\perp/(2\s^2_{ph})$ and $\Delta \spk:=\spk'-\spk_0$. Then
\begin{equation}
\begin{split}
	\int d\spq_\perp d\spp'\rho^{ph}
    (\spk'-\spq_\perp,\spk')\rho^{(1)}(\spp'+\spq_\perp,\spp')&=4\pi \Sigma^2N_h\rho^{ph}(\spk',\spk') e^{\Sigma^2 \mathbf{v}^2_\perp},\\
	\int d\spq_\perp d\spp'q_{\perp}^i
    q_{\perp}^j\rho^{ph}(\spk'-\spq_\perp,\spk')\rho^{(1)}(\spp'+\spq_\perp,\spp')&= 8\pi \Sigma^4N_h\rho^{ph}(\spk',\spk') e^{\Sigma^2 \mathbf{v}^2_\perp}(\delta^{ij}_\perp +2\Sigma^2 v_{\perp}^i v_{\perp}^j),\\
	\int d\spq_\perp d\spp'q_{\perp}^i
    p'^j\rho^{ph}(\spk'-\spq_\perp,\spk')\rho^{(1)}(\spp'+\spq_\perp,\spp')&=-4\pi \Sigma^4N_h\rho^{ph}(\spk',\spk') e^{\Sigma^2 \mathbf{v}^2_\perp}(\delta^{ij}_\perp +2\Sigma^2 v_{\perp}^i v_{\perp}^j -2v^i_\perp p^j_0),
\end{split}
\end{equation}
where $\de_\perp^{ij}:=\delta^{ij}-n'^in'^j$. Substituting these expressions into \eqref{non_pol_dens_out}, \eqref{pol_dens_out}, \eqref{half_pol_dens_out}, we obtain in the case of unpolarized particles that
\begin{equation}\label{pol_dens_out_Gauss}
\begin{split}
	a&=-\frac{2\mu_p^2\Si^4}{\pi M k'_0} N_h \rho^{ph}(\spk',\spk')\im\big[(1+\Si^2\mathbf{v}_\perp^2)e^{\Si^2\mathbf{v}_\perp^2}\big],\\
	b_1&=-\frac{4\mu_p^2\Si^6}{\pi M k'_0n'^2_\perp} N_h
    \rho^{ph}(\spk',\spk')\im\big[(\mathbf{v}_\perp\spd) (\spn'\mathbf{v}_\perp\spd)e^{\Si^2\mathbf{v}_\perp^2}\big],\\
	b_2&=\frac{\mu_p^2\Si^4}{\pi M^2} N_h
    \rho^{ph}(\spk',\spk')\re\big[(\spn'\mathbf{v}_\perp\spp_0)e^{\Si^2\mathbf{v}_\perp^2}\big],\\
	b_3&=-\frac{2\mu_p^2\Si^4}{\pi M k'_0} N_h
    \rho^{ph}(\spk',\spk')\im\big[\big(1+\Si^2\mathbf{v}^2_\perp -\frac{\spd^2_\perp+2\Si^2(\mathbf{v}_\perp\spd)^2}{n'^2_\perp}\big) e^{\Si^2\mathbf{v}_\perp^2}\big].
\end{split}
\end{equation}
As for the case of polarized particles, we have
\begin{equation}\label{non_pol_dens_out_Gauss}
\begin{split}
	a&=-\frac{2\mu_p^2\Si^2}{\pi} N_h\xi^{ph}_2(\spn'\bs\xi^h)
    \rho^{ph}(\spk',\spk')\im e^{\Si^2\mathbf{v}_\perp^2},\\
	b_1&=-\frac{2\mu_p^2\Si^2}{\pi} N_h\xi^{ph}_3(\spn'\bs\xi^h)
    \rho^{ph}(\spk',\spk')\re e^{\Si^2\mathbf{v}_\perp^2},\\
	b_2&=-\frac{2\mu_p^2\Si^2}{\pi} N_h(\spn'\bs\xi^h)
    \rho^{ph}(\spk',\spk')\im e^{\Si^2\mathbf{v}_\perp^2},\\
	b_3&=\frac{2\mu_p^2\Si^2}{\pi} N_h\xi^{ph}_1(\spn'\bs\xi^h)
    \rho^{ph}(\spk',\spk')\re e^{\Si^2\mathbf{v}_\perp^2}.
\end{split}
\end{equation}
In the case of coherent scattering of polarized photons by unpolarized neutrons, we arrive at
\begin{equation}\label{half_pol_dens_out_Gauss}
\begin{split}
	a&=\frac{\mu_p^2\Si^4}{\pi M^2} N_h\xi^{ph}_2
    \rho^{ph}(\spk',\spk')\re\big[(\spn'\mathbf{v}_\perp\spp_0) e^{\Si^2\mathbf{v}_\perp^2}\big],\\
	b_1&=-\frac{\mu_p^2\Si^4}{\pi M^2} N_h\xi^{ph}_3
    \rho^{ph}(\spk',\spk')\im\big[(\spn'\mathbf{v}_\perp\spp_0) e^{\Si^2\mathbf{v}_\perp^2}\big],\\
	b_2&=\frac{\mu_p^2\Si^4}{\pi M^2} N_h
    \rho^{ph}(\spk',\spk')\re\big[(\spn'\mathbf{v}_\perp\spp_0) e^{\Si^2\mathbf{v}_\perp^2}\big],\\
	b_3&=\frac{\mu_p^2\Si^4}{\pi M^2} N_h\xi^{ph}_1
    \rho^{ph}(\spk',\spk') \im\big[(\spn'\mathbf{v}_\perp\spp_0) e^{\Si^2\mathbf{v}_\perp^2}\big],
\end{split}
\end{equation}
where $(\spn'\mathbf{v}_\perp\spp_0)$ is a triple product of vectors.

\begin{figure}[tp]
\centering
\includegraphics*[width=0.45\linewidth]{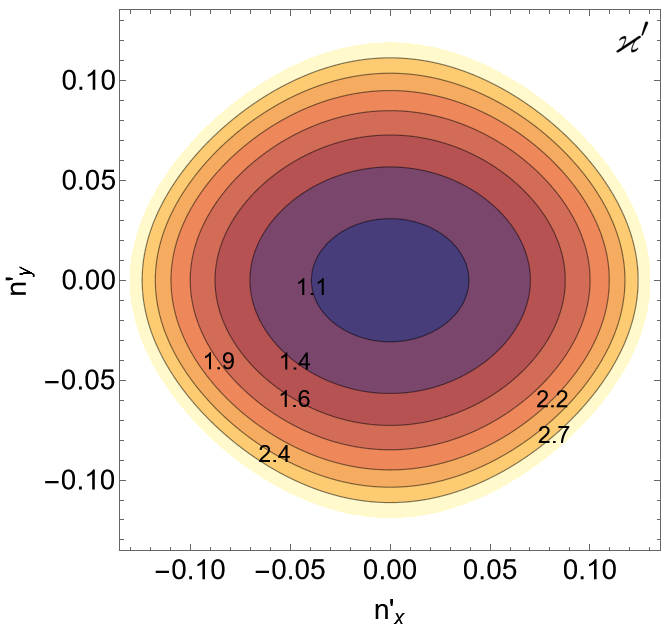}\;
\includegraphics*[width=0.46\linewidth]{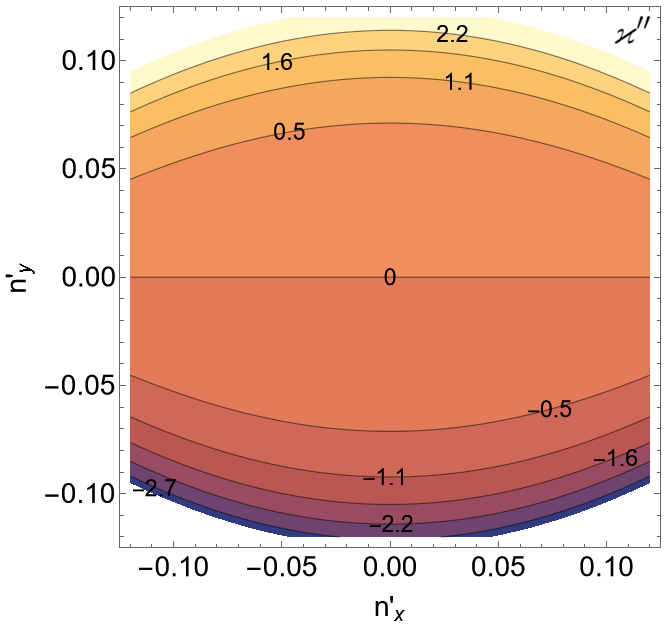}
\caption{{\footnotesize The parameters $\vk'$ and $\vk''$ multiplied by $10^{23}/N_h$ as functions of the direction of emission of the scattered photon $(n'_x,n'_y)$ for coherent elastic scattering of polarized photons by unpolarized neutrons. The parameters $\vk'$ and $\vk''$ are defined in \eqref{de_xi} and determine the evolution of the Stokes vector of scattered photons. The states of photons and neutrons are taken in the form \eqref{Gaussian_states} with the parameters $\spk_0=(0,0,100)$ keV, $\s_{ph}=5$ keV, $\mathbf{g}_{ph}-\mathbf{g}_h=(0,0.2,0)$ keV${}^{-1}$, $\spp_0=(10,0,0)$ keV, $\s_h=5$ keV. }}
\label{kappa_unpolarized_plots}
\end{figure}

Let us find the one-particle density matrix of the scattered photons integrated over the photon energy. The resulting density matrix depends only on the photon emission angle. To this end, we use the standard integral
\begin{equation}
	\int_0^\infty d|\spk'||\spk'|^{\alpha-1} e^{-\frac{(\spk'-\spk_0)^2}{2\s_{ph}^2}}=\s_{ph}^\al \Gamma(\al)
    e^{-\frac{\spk_0^2}{2\s_{ph}^2}+\frac{(\spk_0\spn')^2}{4\s_{ph}^2}}D_{-\al}\Big(-\frac{(\spk_0\spn')}{\s_{ph}}\Big),
\end{equation}
where $D_\al(x)$ is the parabolic cylinder function \cite{GrRy}. Therefore,
\begin{equation}\label{rho_ph_integrtd}
    \tilde{\rho}^{ph}:=\int_0^\infty d|\spk'| |\spk'|^2\rho^{ph}(\spk',\spk')=\frac{N_\ga}{\pi\sqrt{2\pi}} e^{-\frac{\spk_0^2}{2\s_{ph}^2}+\frac{(\spk_0\spn')^2}{4\s_{ph}^2}}D_{-3}\Big(-\frac{(\spk_0\spn')}{\s_{ph}}\Big).
\end{equation}
We will mark all the quantities integrated over $|\spk'|$ by tildes as in \eqref{rho_ph_integrtd}. As a result, in the case of scattering of unpolarized particles, we obtain
\begin{equation}
\begin{split}
	\tilde{a}&=-\frac{\mu_p^2\Si^4}{\pi M\s_{ph}} N_h\tilde{\rho}^{ph}
    \frac{D_{-2}\Big(-\frac{(\spk_0\spn')}{\s_{ph}}\Big)}{D_{-3}\Big(-\frac{(\spk_0\spn')}{\s_{ph}}\Big)} \im\big[(1+\Si^2\mathbf{v}_\perp^2)e^{\Si^2\mathbf{v}_\perp^2}\big],\\
	\tilde{b}_1&=-\frac{2\mu_p^2\Si^6}{\pi M \s_{ph} n'^2_\perp} N_h
    \tilde{\rho}^{ph}
    \frac{D_{-2}\Big(-\frac{(\spk_0\spn')}{\s_{ph}}\Big)}{D_{-3}\Big(-\frac{(\spk_0\spn')}{\s_{ph}}\Big)}
    \im\big[(\mathbf{v}_\perp\spd) (\spn'\mathbf{v}_\perp\spd)e^{\Si^2\mathbf{v}_\perp^2}\big],\\
	\tilde{b}_2&=\frac{\mu_p^2\Si^4}{\pi M^2} N_h
    \tilde{\rho}^{ph} \re\big[(\spn'\mathbf{v}_\perp\spp_0)e^{\Si^2\mathbf{v}_\perp^2}\big],\\
	\tilde{b}_3&=-\frac{\mu_p^2\Si^4}{\pi M \s_{ph}} N_h
    \tilde{\rho}^{ph}
    \frac{D_{-2}\Big(-\frac{(\spk_0\spn')}{\s_{ph}}\Big)}{D_{-3}\Big(-\frac{(\spk_0\spn')}{\s_{ph}}\Big)}
    \im\big[\big(1+\Si^2\mathbf{v}^2_\perp -\frac{\spd^2_\perp+2\Si^2(\mathbf{v}_\perp\spd)^2}{n'^2_\perp}\big) e^{\Si^2\mathbf{v}_\perp^2}\big].
\end{split}
\end{equation}
In the case of polarized particles, we have
\begin{equation}
\begin{split}
	\tilde{a}&=-\frac{2\mu_p^2\Si^2}{\pi} N_h\xi^{ph}_2(\spn'\bs\xi^h)
    \tilde{\rho}^{ph} \im e^{\Si^2\mathbf{v}_\perp^2},\\
	\tilde{b}_1&=-\frac{2\mu_p^2\Si^2}{\pi} N_h\xi^{ph}_3(\spn'\bs\xi^h)
    \tilde{\rho}^{ph} \re e^{\Si^2\mathbf{v}_\perp^2},\\
	\tilde{b}_2&=-\frac{2\mu_p^2\Si^2}{\pi} N_h(\spn'\bs\xi^h)
    \tilde{\rho}^{ph} \im e^{\Si^2\mathbf{v}_\perp^2},\\
	\tilde{b}_3&=\frac{2\mu_p^2\Si^2}{\pi} N_h\xi^{ph}_1(\spn'\bs\xi^h)
    \tilde{\rho}^{ph} \re e^{\Si^2\mathbf{v}_\perp^2}.
\end{split}
\end{equation}
As regards the case of coherent scattering of polarized photons by unpolarized neutrons, we come to
\begin{equation}\label{half_pol_dens_out_Gauss}
\begin{split}
	\tilde{a}&=\frac{\mu_p^2\Si^4}{\pi M^2} N_h\xi^{ph}_2
    \tilde{\rho}^{ph} \re\big[(\spn'\mathbf{v}_\perp\spp_0) e^{\Si^2\mathbf{v}_\perp^2}\big],\\
	\tilde{b}_1&=-\frac{\mu_p^2\Si^4}{\pi M^2} N_h\xi^{ph}_3
    \tilde{\rho}^{ph} \re\big[(\spn'\mathbf{v}_\perp\spp_0) e^{\Si^2\mathbf{v}_\perp^2}\big],\\
	\tilde{b}_2&=\frac{\mu_p^2\Si^4}{\pi M^2} N_h
    \tilde{\rho}^{ph} \re\big[(\spn'\mathbf{v}_\perp\spp_0) e^{\Si^2\mathbf{v}_\perp^2}\big],\\
	\tilde{b}_3&=\frac{\mu_p^2\Si^4}{\pi M^2} N_h\xi^{ph}_1
    \tilde{\rho}^{ph} \re\big[(\spn'\mathbf{v}_\perp\spp_0) e^{\Si^2\mathbf{v}_\perp^2}\big].
\end{split}
\end{equation}
The plots of the parameters $\vk'$ and $\vk''$ determining the evolution of the Stokes vector of polarized scattered photons are presented in Fig. \ref{kappa_polarized_plots} for polarized neutrons and in Fig. \ref{kappa_unpolarized_plots} for unpolarized ones.

\section{Dielectric susceptibility of a neutron wave packet}\label{Dielectric_Suscept_of_Neutr}

A more transparent interpretation can be given to the data of coherent elastic scattering of photons by neutrons obtained in the previous sections if one introduces an effective dielectric susceptibility of a wave packet of a single neutron, $N_h=1$, or a gas of neutrons, $N_h>1$. Then coherent scattering of photons by neutrons, including scattering by a wave packet of a single neutron, can be described as scattering of photons by a medium with a certain dielectric susceptibility. For a single electron wave packet such an interpretation was given in the papers \cite{KazSol2022,AKS2025}.

\subsection{Dielectric susceptibility on the mass-shell}

Recall that we consider a neutral hadron $e_p=0$. Then
\begin{equation}
	\bar{u}^{s'}(\spp') \Ga^{\mu\nu}
    u^s(\spp)=\mu_p^2\big[\delta_{ss'}G_3^{\mu\nu}(\spp,\spp')-(\s_a)_{s's}\tau_{ai}Z^{i\mu\nu}_3(\spp,\spp')\big].
\end{equation}
The amplitude of coherent Compton scattering \eqref{amplitude_coh_Compt} becomes
\begin{equation}
	\Phi_{\la'\la}(\spk',\la';\spk,\la)=
	-2\pi i \frac{\mu^2_p M \bar e^{(\la')}_\mu(\spk') e_{\nu}^{(\la)}(\spk)}{2V\sqrt{k'_0k_0}} \sum_{s,s'}\int
    \frac{d\spp d\spp'}{\sqrt{p_0p'_0}} \delta(p'+k'-p-k) \rho^{(1)}_{ss'}(\spp,\spp')[\delta_{s's}G^{\mu\nu}_3-(\s_a)_{s's}\tau_{ai}Z_3^{i\mu\nu}].
\end{equation}
Representing the delta function as a Fourier transform,
\begin{equation}\label{amplitude_coh_Compt_neutr}
    \Phi_{\la'\la}(\spk',\la';\spk,\la)= - i \frac{\mu^2_p M \bar e^{(\la')}_\mu(\spk') e_{\nu}^{(\la)}(\spk)}{2(2\pi)^3V\sqrt{k'_0k_0}} \int d^4x e^{i(k'-k)x} \sum_{s,s'}\int
    \frac{d\spp d\spp'}{\sqrt{p_0p'_0}} e^{i(p'-p)x} \rho^{(1)}_{ss'}(\spp,\spp')[\delta_{s's}G^{\mu\nu}_3-(\s_a)_{s's}\tau_{ai}Z_3^{i\mu\nu}],
\end{equation}
and comparing this expression with formula (33) of \cite{KazSol2023}, we extract the Weyl symbol of the operator of dielectric susceptibility
\begin{equation}
    \chi^{ij}(x,k_c)= -\mu^2_p M \sum_{s,s'}\int
    \frac{d\spp d\spp'}{(2\pi)^3k'_0k_0} e^{i(p'-p)x} \frac{\rho^{(1)}_{ss'}(\spp,\spp')}{\sqrt{p_0p'_0}}[\delta_{s's}G^{ij}_3-(\s_a)_{s's}\tau_{ak}Z_3^{kij}],
\end{equation}
where
\begin{equation}
    k_0=|\spk_c-\spq/2|,\qquad k'_0=|\spk_c+\spq/2|.
\end{equation}
This expression formally coincides with expression (56) of \cite{AKS2025} if one introduces the time-dependent one-particle density matrix,
\begin{equation}
    \rho^{(1)}_{ss'}(x^0;\spp,\spp')=e^{i(p'_0-p_0)x^0}\rho^{(1)}_{ss'}(\spp,\spp'),
\end{equation}
and passes to the new integration variables $\spp_c$ and $\spq$. Then
\begin{equation}\label{chi_gen}
    \chi^{ij}(x,k_c)= -\mu^2_p M \sum_{s,s'} \int
    \frac{d \spp_c d \spq }{(2\pi)^3 k'_0k_0} e^{i\spq\spx} \frac{\rho^{(1)}_{ss'}(x^0;\spp_c+\spq/2,\spp_c-\spq/2)}{\sqrt{p_0p'_0}} [\delta_{s's}G^{ij}_3-(\s_a)_{s's}\tau_{ak}Z_3^{kij}].
\end{equation}
In comparing this expression with formula (56) of \cite{AKS2025}, one should bear in mind that a dielectric susceptibility differs from a polarization operator by the factor $1/(k'_0k_0)$.

Consider the particular cases of the general expression \eqref{chi_gen}. Let us introduce the Wigner function for the one-particle density matrix of neutrons
\begin{equation}\label{WignerDef}
	\rho^{(1)}_{ss'}(x,\spp_c)=\int
    \frac{d\spq}{(2\pi)^3}e^{i\spq\spx}\rho^{(1)}_{ss'}(x^0;\spp_c+\spq/2,\spp_c-\spq/2),
\end{equation}
and decompose it in the basis of $\s$-matrices,
\begin{equation}\label{WignerDef_sigma}
	\rho^{(1)}_{ss'}(x,\spp_c)=\frac12 \rho^{(1)}(x,\spp_c) \big[\delta_{ss'}+\xi_a(x,\spp_c)(\s_a)_{ss'}\big].
\end{equation}
Then for the polarized neutrons, i.e., for $|\bs\xi^h|\gg|\spq|/M$, it follows from \eqref{chi_gen} and \eqref{ee_G3_Z3} in the small recoil limit that
\begin{equation}\label{chi_xi_non0}
    \chi^{ij}(x,k_c)=-\frac{2i\mu_p^2 M}{(k_c^0)^2} \e^{\la ij\rho} k^c_\rho\int\frac{d\spp_c}{p_0^c}   \rho^{(1)}(x,\spp_c)s^h_\la(x,\spp_c),
\end{equation}
where $s^\mu_h(x,\spp_c)$ is constructed as in \eqref{s_h_mu_defn} with the aid of $\xi^i(x,\spp_c)=\tau_a^i\xi_a(x,\spp_c)$ and $\spp_c$. In evaluating the dielectric susceptibility \eqref{chi_xi_non0}, it has been taken into account that the leading contribution comes from the tensor $Z^{kij}_3$, where it is implied that it should be contracted with the polarization vectors for the dielectric susceptibility on the mass-shell to be found (see \eqref{amplitude_coh_Compt_neutr} and \eqref{ee_G3_Z3}). It is seen that, under the  specified conditions, the wave packet of a polarized neutron behaves as a gyrotropic medium in a coherent Compton process (see, e.g., Sec. 101 of \cite{LandLifshECM}).

If the neutrons are unpolarized, viz., $|\bs\xi^h|\ll|\spq|/M$, and the average modulus of momenta in the one-particle density matrix of neutrons satisfies $|\spp|/M\gg|\spq|/k^0_c$, then we obtain from \eqref{chi_gen} and \eqref{ee_G3_Z3} in the small recoil limit that
\begin{equation}\label{chi_xi0}
    \chi_{ij}(x,k_c)=\frac{i\mu_p^2}{k_c^0}  \frac{\partial}{\partial x^{[i}} \int\frac{d\spp_c}{p_0^c(M+p_0^c)} p^c_{j]}  \rho^{(1)}(x,\spp_c)\approx \frac{i\mu_p^2}{k_c^0M(M+p_0^c)} \frac{\partial}{\partial x^{[i}} \rho^{(1)}(x)p^c_{j]},
\end{equation}
where it has been assumed in the last approximate equality that the dispersion of momenta in the one-particle density matrix of neutrons is sufficiently small so that the $4$-vector $p_c^\mu$ is replaced by its average value, and
\begin{equation}
    \rho^{(1)}(x):=M\int\frac{d\spp_c}{p_0^c}\rho^{(1)}(x,\spp_c).
\end{equation}
It has been taken into account in evaluating the expression for the dielectric susceptibility \eqref{chi_xi0} that, in the case at hand, the leading contribution stems from the linear in $q^\mu$ terms in the tensor $G^{ij}_3$ (see \eqref{ee_G3_Z3}). The dielectric susceptibility of the form \eqref{chi_xi0} arises in describing dynamo-optical phenomena in inhomogeneously moving fluids (see, e.g., Sec. 102 of \cite{LandLifshECM}). In our case, it describes a gyrotropic medium.

In the case when the neutrons are unpolarized and the average modulus of momenta in the one-particle density matrix is such that $|\spp|/M\ll|\spq|/k^0_c$, then the leading contribution to the dielectric susceptibility \eqref{chi_gen} is given by the tensor $G^{ij}_3$, where $p_c^\mu=(M,0,0,0)$. Therefore we deduce from \eqref{chi_gen} and \eqref{ee_G3_Z3} that
\begin{equation}\label{chi_xi00}
    \chi_{ij}(x,k_c)=\frac{\mu_p^2}{M(k_c^0)^2} (\de_{ij}\De-\partial_i\partial_j)\rho^{(1)}(x).
\end{equation}
It is clear that the resulting susceptibility tensor is divergenceless in this case. Besides, its dependence on the photon energy $k_c^0$ is the same as for the plasma dielectric susceptibility.

\subsection{Off-shell polarization operator}\label{Off-shell_Polarization_Operator}

Let us find the one-loop correction to the polarization operator of a photon in the presence of a wave packet of a neutron or neutrons with one-particle density matrix \eqref{density_oper}. The evaluation of such a polarization operator is carried out along the lines of the calculations of polarization operator of a photon in the presence of an electron wave packet \cite{AKS2025}. In the leading order of perturbation theory and for the small photon momenta in comparison with the rest energy of $\pi$-meson in the neutron rest frame, one can neglect the vacuum contribution to the one-loop polarization operator. Then the Weyl symbol of the polarization operator of a photon in the presence of neutrons with the one-particle density matrix \eqref{density_oper} can be cast into the form (see the details of calculations in \cite{AKS2025}, cf. \eqref{chi_gen})
\begin{equation}\label{polariz_oper}
\begin{split}
    \Pi^{\mu\nu}(x,k_c)=&-\mu_p^2 M\sum_{s,s'} \int\frac{d\spp_c d\spq}{(2\pi)^3} e^{i\spq\spx}\frac{\rho^{(1)}_{ss'}(x^0;\spp_c+\spq/2,\spp_c-\spq/2)}{\sqrt{p_0p_0'}} \times\\
    &\times \big[\de_{s's}G_3^{\mu\nu}(\spp_c+\spq/2,\spp_c-\spq/2) -(\s_a)_{s's}\tau_{ak} Z_3^{k\mu\nu}(\spp_c+\spq/2,\spp_c-\spq/2) \big].
\end{split}
\end{equation}
As long as this expression is valid for both a single neutron and a gas of neutrons with the one-particle density matrix \eqref{density_oper}, in the present section we will use the terms the wave packet of a neutron and the one-particle density matrix of neutrons interchangeably.

Let us find the approximate expression for \eqref{polariz_oper} in the short wavelength approximation, i.e., when the scale of variations of the external electromagnetic field is much smaller than the typical scale of variations of the one-particle density matrix of neutrons in the $x$-space. As is seen from \eqref{polariz_oper}, $|\spq|\approx\s_h$, where $\s_h$ is the typical scale of variations of the one-particle density matrix in the momentum space. Thus, in the short wavelength approximation,
\begin{equation}\label{short_wave_approx}
    |k^\mu|\gg |q^\mu|.
\end{equation}
Furthermore, we suppose that the one-particle density matrix of neutrons in sufficiently narrow in the momentum space,
\begin{equation}
    |\spp_c|\approx|\spp_c^0|\gg|\spq|,
\end{equation}
where $\spp_0$ is the average value of momentum in the one-particle density matrix of neutrons. In this case, employing expressions  \eqref{G3_Z3_off-shell} for the tensors $G_3^{\mu\nu}$ and $Z_3^{\mu\nu}$ in the small recoil limit, we have
\begin{equation}
    \Pi^{\mu\nu}(x,k_c)=-\mu_p^2 M \int\frac{d\spp_c}{p_c^0} \rho^{(1)}(x;\spp_c)
    \big[G_3^{\mu\nu}(\spp_c,\spp_c) -\xi^h_{k}(x,\spp_c) Z_3^{k\mu\nu}(\spp_c,\spp_c) \big],
\end{equation}
where the Wigner function has been introduced in the form \eqref{WignerDef_sigma}. Assume additionally that the one-particle density matrix of neutrons describes the state with a certain polarization, viz., one can put $\spp_c\rightarrow\spp^0_c$ in $\xi^h_{k}(x,\spp_c)$. Then, for a one-particle density matrix of neutrons that is narrow in momentum space, we deduce
\begin{equation}\label{polariz_oper_shrt_wl}
\begin{split}
    \Pi^{\mu\nu}(x,k)=&\frac{\mu_p^2(x)}{(kp)^2-k^4/4} \Big[\frac{k^4}{M} (M^2\eta^{\mu\nu} -p^\mu p^\nu) +\frac{k^2}{M}(kp) k^{(\mu} p^{\nu)} -(M^2k^2+(kp)^2)\frac{k^\mu k^\nu}{M}+\\
    &+2i(kp)(ks_h) \e^{\mu\nu\rho\s}k_\rho p_\s -i\frac{k^4}{2} \e^{\mu\nu\rho\s} k_\rho s^h_\s \Big],
\end{split}
\end{equation}
where $\mu^2_p(x):=\mu^2_p\rho^{(1)}(x)$ is the density of a magnetic moment squared. Henceforth, for short, the index $0$ at the average momentum $p^\mu_0$ is omitted. The terms in the square brackets that do not depend on $s_h^\mu$ can be written as
\begin{equation}
    \frac1{M} (k^2\eta^\mu_{\ \mu'}-k^\mu k_{\mu'}) (M^2\eta^{\mu'\nu'}-p^{\mu'}p^{\nu'}) (k^2\eta_{\nu'}^{\ \nu}-k_{\nu'} k^{\nu}).
\end{equation}
Then it is clear that the polarization operator \eqref{polariz_oper_shrt_wl} obeys the Ward identity $\Pi^{\mu\nu}(x,k)k_\nu=0$. It is seen that, as in the case of an electron wave packet \cite{AKS2025}, the polarization operator of a photon in the presence of a wave packet of a neutron is singular when
\begin{equation}\label{plasmon_disp_law}
    (kp)^2-k^4/4=0,
\end{equation}
that manifests the presence of quasiparticles in the theory -- plasmons -- with the dispersion law determined by equation \eqref{plasmon_disp_law}. These quasiparticles exist even in the case of a wave packet of a single neutron. The plasmon modes hybridize with the electromagnetic modes and result in plasmon-polariton modes that are the solutions of the effective Maxwell equations in the presence of a neutron wave packet or a gas of neutrons with the corresponding one-particle density matrix.

In order to obtain the solutions of the effective Maxwell equations, we suppose that $\mu^2_p(x)$ is approximately constant, i.e., $\rho^{(1)}(x)$ is a slowly varying function of $x$ on the typical scale of variations of the electromagnetic field. On performing the Fourier transform,
\begin{equation}
    A_\mu(x)=\int\frac{d^4k}{(2\pi)^4}e^{-ikx}A_\mu(k),
\end{equation}
the effective Maxwell equations are reduced to the matrix equation
\begin{equation}\label{Maxwell_eqs}
\begin{split}
    \Big\{-k^2\eta^{\mu\nu}+k^\mu k^\nu &+\frac{\mu_p^2(x)}{(kp)^2-k^4/4}\Big[\frac{k^4}{M} (M^2\eta^{\mu\nu} -p^\mu p^\nu) +\frac{k^2}{M}(kp) k^{(\mu} p^{\nu)}-\\
    &-(M^2k^2+(kp)^2)\frac{k^\mu k^\nu}{M}
    +2i(kp)(ks_h) \e^{\mu\nu\rho\s}k_\rho p_\s -i\frac{k^4}{2} \e^{\mu\nu\rho\s} k_\rho s^h_\s\Big] \Big\}A_\nu(k)=0.
\end{split}
\end{equation}
These equations are explicitly Lorentz-covariant. Therefore, it is convenient to pass to the rest frame of the neutron, where $p^\mu=(M,0,0,0)$, to find the solution of equations \eqref{Maxwell_eqs} in this frame, and then to revert to the original reference frame by the Lorentz transform. In order to simplify the notation, we arrange that the $4$-momentum of a photon is measured in the rest energies of a neutron, $k^\mu\rightarrow Mk^\mu$, and introduce the dimensionless quantity
\begin{equation}\label{mus}
    \tilde{\mu}^2:=\mu^2_p(x)/M.
\end{equation}
Notice that in virtue of the approximations made in derivation of the polarization operator only the domain of photon momenta $|k^\mu|\ll1$ in the neutron rest frame makes sense.

\begin{figure}[tp]
\centering
\includegraphics*[width=0.45\linewidth]{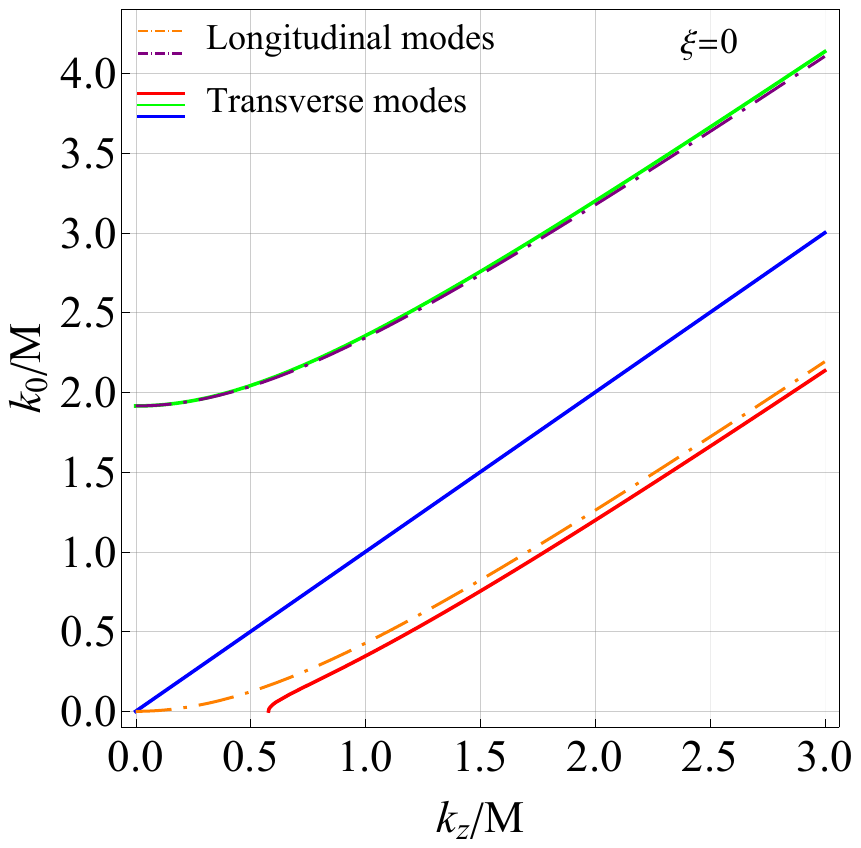}\;
\includegraphics*[width=0.45\linewidth]{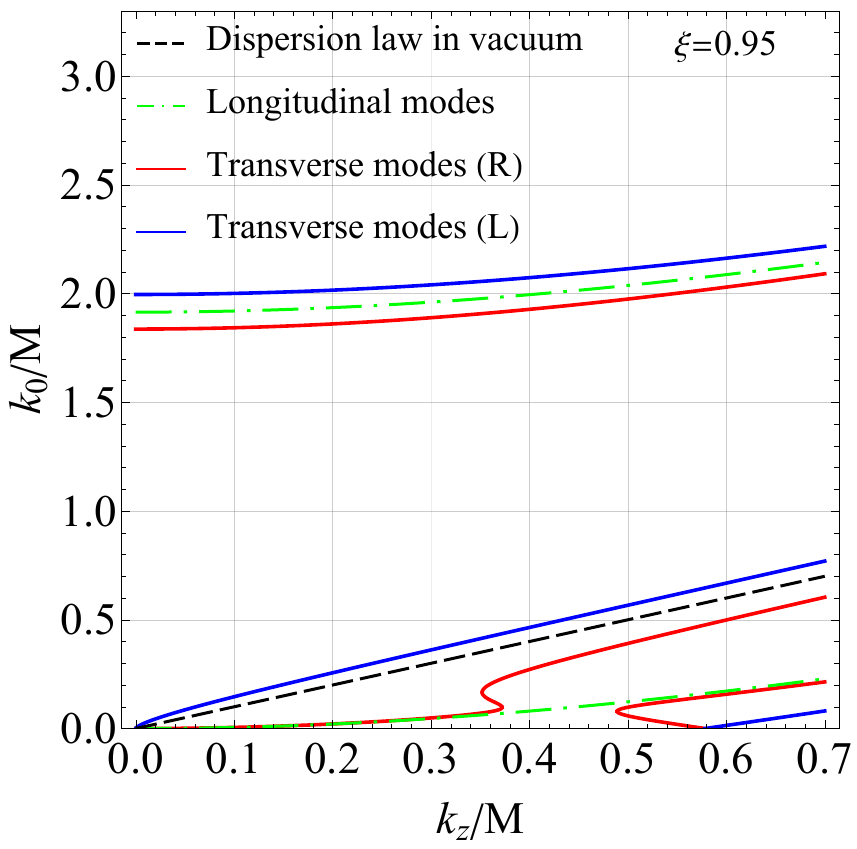}
\caption{{\footnotesize Left panel: the dispersion laws of plasmon-polariton modes on a single unpolarized neutron in its rest frame. The parameter \eqref{mus} is $\tilde{\mu}^2=0.08$. Such a value of this parameter is taken to show the main peculiarities of the dispersion laws. The instability of the transverse mode at $|k_z|<2\tilde{\mu}$ is clearly seen. Right panel: the same as on the left panel but for a single polarized neutron with $\xi=0.95$ and $\spk\parallel\mathbf{s}_h$. It is clear that the transverse modes possess instabilities. The estimates for the points where the group velocity is infinity are given in \eqref{sing_points}. It is also seen that the two transverse modes are tachyonic even at large momenta. The transverse mode, which dispersion curve is slightly above the free dispersion law, is tachyonic for any value of the momentum.}}
\label{dispersion_laws_plots}
\end{figure}

Let us consider, at first, the case of an unpolarized wave packet of a neutron, $s^\mu_h=0$. In this case, rotating the reference frame, one can always bring the photon momentum to the form $k^\mu=(k_0,0,0,k_z)$. Then there are eight independent physical solutions to the effective Maxwell equations \eqref{Maxwell_eqs}. The two independent modes take the form
\begin{equation}\label{longitud_modes_s0}
    A^\mu(k)=(k_z,0,0,k_0),
\end{equation}
with the dispersion laws
\begin{equation}\label{longitud_modes_disp_s0}
    k_0=\sqrt{1-\tilde{\mu}^2+k_z^2} \pm\sqrt{1-\tilde{\mu}^2},
\end{equation}
where it is assumed that $\tilde{\mu}^2<1$. The longitudinal modes \eqref{longitud_modes_s0} describe oscillations of a homogeneous electric field along the $z$ axis. The six independent transverse modes are written as
\begin{equation}\label{transverse_modes_s0}
    A^\mu(k)=(0,\al,\be,0),
\end{equation}
where $\al$ and $\be$ are some complex constants independent of the momentum $k^\mu$. The two independent modes of the form \eqref{transverse_modes_s0} have the free dispersion law
\begin{equation}
    k_0=|k_z|,
\end{equation}
whereas the rest transverse modes \eqref{transverse_modes_s0} possess the dispersion laws
\begin{equation}\label{transverse_modes_disp_s0}
    k_0=\big[\big(\sqrt{(1-\tilde{\mu}^2)^2+k_z^2} \pm1-\tilde{\mu}^2 \big)\big(\sqrt{(1-\tilde{\mu}^2)^2+k_z^2} \pm1+\tilde{\mu}^2 \big) \big]^{1/2},
\end{equation}
where the signs in the parentheses are agreed. As is seen, the plasmon-polariton mode corresponding to the minus sign in \eqref{transverse_modes_disp_s0} becomes unstable for
\begin{equation}\label{instab_region}
    |k_z|<2\tilde{\mu},
\end{equation}
where by definition $\tilde{\mu}>0$. For $|k_z|\geqslant2\tilde{\mu}$ close to the point $|k_z|=2\tilde{\mu}$, the group velocity of this plasmon-polariton mode is greater than unity (see Fig. \ref{dispersion_laws_plots}), i.e., this mode is tachyonic in this region of momenta, that corroborates once more the presence of instability (for a recent discussion of the relation between stability and causality see, e.g., \cite{Gavassino2024,Hoult2024,Wang2024}).

As a result, there is an exponential growth of the transverse mode in the momentum domain \eqref{instab_region} that can be interpreted as the spontaneous onset of magnetization. To put it another way, if the condition \eqref{instab_region} is satisfied and the conditions that was assumed in deriving expression \eqref{polariz_oper_shrt_wl} for the polarization operator are also fulfilled, the state of neutrons is ferromagnetic. Let us estimate the values of parameters of the one-particle density matrix of neutrons that give rise to instability. Combining \eqref{short_wave_approx} and \eqref{instab_region} and restoring $M$, we obtain that the instability takes place when
\begin{equation}\label{instab_region_est1}
    \s_h^2 \ll 4M\mu^2_p(x).
\end{equation}
For the estimates, we suppose that $\rho^{(1)}(x)\sim N_h \s_h^3$. Hence, taking into account the value of the neutron anomalous magnetic moment, the estimate \eqref{instab_region_est1} turns into
\begin{equation}\label{instab_region_est2}
    1\ll16\pi\al N_h\s_h/M.
\end{equation}
In particular, it follows from this estimate that the instability is not realized for a single neutron, $N_h=1$, provided the wave packet of a neutron is not localized on the scale much smaller than its Compton wavelength. Nevertheless, for sufficiently large $N_h$, the estimate can be satisfied. Assuming that the distribution over momenta in the one-particle density matrix of neutrons is approximately Gaussian (the nondegenerate Fermi gas), we introduce the effective temperature
\begin{equation}
    T:=\s_h^2/M.
\end{equation}
Introduce also the nuclear concentration (see, e.g., \cite{HaensPtoYak2007})
\begin{equation}
    \rho_0=\frac{3}{4\pi r_0^3},\qquad r_0\approx6/M.
\end{equation}
Then the condition \eqref{instab_region_est1} is equivalent to
\begin{equation}\label{instab_region_est3}
    \rho^{(1)} \gg \frac{18}{\al}\frac{T}{M}\rho_0.
\end{equation}
For example, for ultracold neutrons \cite{Ignatovich1996,Serebrov2011,Pokotilovski2018,Lauss2021} with $\s_h=1$ eV, $T=1.06 $ neV, the quantity on the right-hand side  equals
\begin{equation}
    \frac{18}{\al}\frac{T}{M}\rho_0\approx 3\times 10^{23}\,\text{cm}^{-3},
\end{equation}
that corresponds to the mass density of order of one half of the water mass density at normal conditions. Usually, the ultracold neutrons have temperatures of order $T\approx 100$ neV and so the required density is two orders larger than that of a water. It is clear that at such densities the short-range nuclear interaction between neutrons is negligible.

Now we consider the case of the polarized state of a neutron. Let us find, at first, the plasmon-polariton modes propagating along the spin vector, i.e., with $\spk\parallel\mathbf{s}_h$. Setting the $z$ axis along the spin vector, we obtain
\begin{equation}\label{spin_vector}
    s^\mu_h=(0,0,0,\xi),\qquad |\xi|\leqslant1.
\end{equation}
In this case, there are also eight independent plasmon-polariton modes. The two longitudinal modes are the same as in the case of an unpolarized neutron \eqref{longitud_modes_s0}, \eqref{longitud_modes_disp_s0}. The six transverse modes,
\begin{equation}\label{transverse_modes_sn0}
    A^\mu(k)=(0,1,\pm i,0),
\end{equation}
possess the dispersion laws specified by the sixth-order polynomial equation with respect to $k_0$,
\begin{equation}\label{transverse_modes_disp_sn0}
    (k_0^2-k_z^2)((k_0^2-k_z^2)^2 -4k_0^2)+\tilde{\mu}^2\big[4(k_0^2-k_z^2)^2 \pm 2\xi k_0((k_0^2-k_z^2)^2 -4k_z^2) \big]=0.
\end{equation}
The signs ``$\pm$'' in \eqref{transverse_modes_sn0} and \eqref{transverse_modes_disp_sn0} are agreed. As is seen, the dispersion law has the symmetry
\begin{equation}\label{disp_law_sym}
    k_0\rightarrow-k_0,\qquad \xi\rightarrow-\xi.
\end{equation}
Therefore, there are only six branches of the dispersion law with positive $k_0$ for both signs in \eqref{transverse_modes_sn0}. Moreover, the dispersion law is invariant under $k_z\rightarrow-k_z$. Equation \eqref{transverse_modes_disp_sn0} is a third order polynomial equation with respect to $k_z^2$ and its solution can be found explicitly. However, it is rather huge and we do not write out it here. In the particular case $|\xi|\rightarrow1$, equation \eqref{transverse_modes_disp_sn0} simplifies and has the solutions
\begin{equation}\label{disp_law_s1}
    k_z^2=k_0^2 \pm 2\xi k_0,\qquad k_z^2=k_0^2+2-u(\pm\xi k_0+2) \pm\sqrt{\big(k_0^2+2-u(\pm\xi k_0+2)\big)^2- k_0^3( k_0 \mp 2\xi u)},
\end{equation}
where the signs at $\xi$ are agreed with \eqref{transverse_modes_sn0} while the different signs in front of the square root characterize the different branches of the dispersion law and are independent of the signs in \eqref{transverse_modes_sn0}. Also, for brevity, we have introduced the notation $u:=1-\tilde{\mu}^2$. Notice that the first expression in \eqref{disp_law_s1} coincides with the dispersion law of plasmons \eqref{plasmon_disp_law}. As a rule, $\tilde\mu^2$ is a tiny positive quantity. Therefore, for $|k_0|\gg\tilde{\mu}^2$ and $|k_z|\gg\tilde{\mu}^2$, it makes sense to expand the dispersion laws determined by equation \eqref{transverse_modes_disp_sn0} with respect to $\tilde{\mu}^2$ and to keep only the leading contributions. Then, for the right-handed mode \eqref{transverse_modes_sn0}, we arrive at
\begin{equation}
    k_0=-1\pm\sqrt{1-2(1-\xi)\tilde{\mu}^2+k_z^2},\qquad k_0=1\pm\sqrt{1-2(1+\xi)\tilde{\mu}^2+k_z^2},\qquad k_0=-\xi\tilde{\mu}^2 \pm\sqrt{\xi^2\tilde{\mu}^4 +k_z^2}.
\end{equation}
The corresponding expressions for the left-handed mode \eqref{transverse_modes_sn0} are obtained by the replacement $\xi\rightarrow-\xi$. The plots of the dispersion laws of plasmon-polaritons are given in Fig. \ref{dispersion_laws_plots}. As in the case of an unpolarized state of neutrons, there exists instability at certain values of momenta $k_z$ and, consequently, there are the domains of momenta where the dispersion laws possess the group velocity larger than the speed of light. It is not difficult to deduce the estimates for the points where the branches of the dispersion law have an infinite group velocity. There are three such points for nonzero $|k_z|$:
\begin{equation}\label{sing_points}
    |k_z|\approx3^{3/2}\tilde{\mu}^2,\; k_0\approx \pm 3\xi\tilde{\mu}^2;\qquad |k_z|=\tilde{\mu} \big[2(1\pm\sqrt{1-\xi^2})\big]^{1/2},\; k_0\approx \pm\xi\tilde{\mu}^2.
\end{equation}
Furthermore, for $k_z\rightarrow0$, one of the branches of the dispersion law behaves as
\begin{equation}\label{sing_disp_law}
    k_0\approx\Big(\mp\frac{2\tilde{\mu}^2\xi}{1-\tilde{\mu}^2}\Big)^{1/3} |k_z|^{2/3},
\end{equation}
i.e., it possesses an infinite group velocity for $k_z\rightarrow0$. The signs ``$\pm$'' in \eqref{sing_points}, \eqref{sing_disp_law} are agreed with \eqref{transverse_modes_sn0}. The position of the singular point with the greatest $|k_z|$, which is given by the second expression in \eqref{sing_points} with the plus sign, provides the estimate for the value of momentum below which the plasmon-polariton modes are unstable. For $\xi=0$, this estimate coincides with \eqref{instab_region}. Notice that a tachyonic behavior of some branches of the dispersion law for plasmon-polariton modes on polarized neutrons is preserved even at large momenta (see Fig. \ref{dispersion_laws_plots}). For example, it follows from \eqref{disp_law_s1} that the dispersion laws specified by the second equality in \eqref{disp_law_s1} have the asymptotics
\begin{equation}\label{disp_law_asympt}
    k_0=k_z\pm\tilde{\mu}^2 -\frac{(3-\tilde{\mu}^2)\tilde{\mu}^4}{2(1+\tilde{\mu}^2) k_z}+O(1/k_z^2),\qquad k_0=k_z\pm 1 +\frac{\frac{2}{1+\tilde{\mu}^2}-\frac32}{k_z}+O(1/k_z^2),
\end{equation}
where both the right-handed and left-handed modes \eqref{transverse_modes_sn0} are considered. As long as $\tilde{\mu}^2<1$, the former branches of the dispersion law \eqref{disp_law_asympt} are slightly tachyonic that reflects instability of the system. It is one of these tachyonic branches of the dispersion law that has the infrared asymptotics \eqref{sing_disp_law}.

Let us consider the general case. Without loss of generality, we suppose that the spin vector takes the form \eqref{spin_vector} and the momentum vector has the components
\begin{equation}
    k^\mu=(k_0,0,k_\perp,k_z).
\end{equation}
Introduce the notation
\begin{equation}
\begin{gathered}
    h_\parallel := k^4-4k_0^2(1-\tilde{\mu}^2),\qquad h^\pm_\perp:=k^2(k^4-4k_0^2+4\tilde{\mu}^2 k^2) \pm2k_0\xi\tilde{\mu}^2(k^4-4\spk^2),\\
    h_\perp=h^\pm_\perp|_{\xi=0}=k^2(k^4-4k_0^2+4\tilde{\mu}^2 k^2).
\end{gathered}
\end{equation}
Then the eight plasmon-polariton modes can be cast into the form (cf. (85) of \cite{AKS2025})
\begin{equation}\label{plasmon_polariton_modes}
    A^\mu(k)=\big(2\xi\tilde{\mu}^2k_\perp k^2h_\perp,ih_\perp h_\parallel,2\xi\tilde{\mu}^2k_0[k^2 h_\perp +4k_z^2(1-\tilde{\mu}^2)(4k_0^2-k^4)],-8\xi\tilde{\mu}^2(1-\tilde{\mu}^2)k_0 k_\perp k_z(4k_0^2-k^4) \big),
\end{equation}
where $k_0$ is found from the dispersion law determined by the equation
\begin{equation}\label{plasmon_polariton_disp_law}
    h_\parallel h^+_\perp h^-_\perp +4\xi^2\tilde{\mu}^4 k_\perp^2(k^4-4k_0^2)\big[k^4(k^4-4\spk^2) -4(1-\tilde{\mu}^2)(k^6+k_0^2(k^4-4\spk^2))\big]=0.
\end{equation}
This is a sixteenth-order polynomial equation with respect to $k_0$. This equation obeys the symmetry \eqref{disp_law_sym} and is also invariant under the replacements $k_z\rightarrow-k_z$ and $k_\perp\rightarrow-k_\perp$. Notice that the equations $h_\parallel=0$ and $h_\perp=0$ with $k_\perp=0$ give rise to the dispersion laws of longitudinal and transverse plasmon-polariton modes, respectively, in the case of unpolarized neutrons, $\xi=0$.

\subsection{Polarization operator in the infrared limit}

In this section we obtain the infrared limit of the photon polarization operator in the presence of a wave packet of a spin polarized neutron. In the infrared limit, the wavelength of an external electromagnetic field is supposed to be much larger than the typical size of the neutron wave packet in the coordinate space. We also assume that the neutron wave packet is sufficiently narrow in the momentum space. Since the general expression for the polarization operator \eqref{polariz_oper} is determined by the one-particle density matrix, the approximate expression that will be derived below is valid not only for a single neutron but also for a gas of free neutrons with spin polarized one-particle density matrix. The support of the Wigner function for this density matrix in the coordinate space must be much smaller than the wavelength of the external electromagnetic field and this density matrix must be narrow in the momentum space. The derivation procedure of the polarization operator in the infrared limit is analogous to the procedure presented in \cite{AKS2025} for a wave packet of a single electron.

The above assumptions boil down to the estimates
\begin{equation}\label{IR_estimates}
    |k^\mu|\ll |p^\mu|,\qquad |q^\mu|\ll|p^\mu|,
\end{equation}
where $p^\mu$ is the typical value of the momentum in the wave packet and
\begin{equation}
    \rho_{ss'}(\spp,\spp')\approx \rho_{ss'}(\spp_c,\spp_c)e^{-i\spq\spx_0},
\end{equation}
where $\spx_0$ is the position of the center of the neutron wave packet at the instant of time $t=0$ and $\rho_{ss'}(\spp_c,\spp_c)$ is supposed to be concentrated near the momentum $\spp_c^0$. For a spin polarized neutron wave packet, under the assumptions \eqref{IR_estimates}, the contribution of the tensor $G_3^{\mu\nu}$ to the polarization operator \eqref{polariz_oper} turns out to be suppressed as compared to the contribution of $Z^{i\mu\nu}_3$. Then we obtain in the infrared limit
\begin{equation}\label{pi_munu_defn}
    \xi_k(\spp_c,\spp_c) Z^{k\mu\nu}\approx\pi^{\mu\nu}(\spk',\spk)=ip^c_\la \Big[\frac{s_h^\mu\e^{\nu\rho\s\la}k'_\rho k_\s +(s_hk')\e^{\mu\nu\rho\la}k_\rho }{(p_ck')} +\frac{s_h^\nu\e^{\mu\rho\s\la}k'_\rho k_\s +(s_hk)\e^{\mu\nu\rho\la}k'_\rho }{(p_ck)}\Big].
\end{equation}
This expression satisfies the Ward identities
\begin{equation}
    k'_\mu\pi^{\mu\nu}(k',k)=\pi^{\mu\nu}(k',k)k_\nu=0.
\end{equation}
Representing the one-particle density matrix of neutrons in the form \eqref{dens_mat_ini}, the contribution of the polarization operator to the effective Maxwell equations in the momentum representation becomes
\begin{equation}
    \int \frac{d^4k}{(2\pi)^4}\Pi^{\mu\nu}(k',k) A_\nu(k)\approx \mu^2_p M\int\frac{d\spp_c d\spq}{(2\pi)^3p_c^0} \rho^{(1)}(\spp_c,\spp_c)e^{-i\spq\spx_0} \pi^{\mu\nu}(k',k'-q) A_\nu(k'-q).
\end{equation}
As long as $\rho^{(1)}(\spp_c,\spp_c)$ is concentrated near $\spp_c^0$ and the normalization condition \eqref{normalization_dens_matr} is fulfilled, we have
\begin{equation}\label{Pi_A_IR}
    \int \frac{d^4k}{(2\pi)^4}\Pi^{\mu\nu}(k',k) A_\nu(k)\approx \frac{\mu_p^2N_h M}{p^0_c} \int \frac{d\spq}{(2\pi)^3} e^{-i\spq\spx_0} \pi^{\mu\nu}(k',k'-q)\int d^4x e^{i(k'-q)x}A_\nu(x),
\end{equation}
where, in the last expression, one ought to replace $\spp_c\rightarrow\spp_c^0$. In what follows, for brevity, we do not write the index $0$ at $\spp_c^0$ and the quantities involving $\spp_c^0$.

Expression \eqref{Pi_A_IR} can be rewritten as
\begin{equation}
    \int \frac{d^4k}{(2\pi)^4}\Pi^{\mu\nu}(k',k) A_\nu(k)\approx \frac{\mu_p^2N_h M}{p^0_c}\int d^4x e^{ik'x} \pi^{\mu\nu}\Big(i\frac{\partial}{\partial x},i\frac{\partial}{\partial x}\Big|_{A}\Big) A_\nu(x) \de(\spx-\spx_0-\bs\be_c x^0),
\end{equation}
where $\partial/\partial x|_{A}$ denotes the derivative that acts on the variable $x^\mu$ only in the potential $A_\nu(x)$. Introduce the trajectory of the center of the one-particle density matrix of neutrons in the coordinate space
\begin{equation}
    x^\mu(\tau)=x_0^\mu +\frac{p_c^\mu}{M}\tau,
\end{equation}
where $\tau$ is the natural parameter. Bearing in mind that
\begin{equation}
    \de(\spx-\spx_0-\bs\be_c x^0)=\int d\tau\frac{p_c^0}{M} \de(x-x(\tau)),
\end{equation}
we bring the effective Maxwell equations in the coordinate representation into the form
\begin{equation}\label{Max_eqns_eff_IR}
    (\Box\eta^{\mu\nu}-\partial^\mu\partial^\nu)A_\nu(x) +\mu_p^2N_h \int d\tau \pi^{\mu\nu}\Big(i\frac{\partial}{\partial x},i\frac{\partial}{\partial x}\Big|_{A}\Big) \de(x-x(\tau)) A_\nu(x)=0.
\end{equation}
As in the case of the wave packet of a single electron, the polarization operator $\pi^{\mu\nu}$ possesses singularities (see \eqref{pi_munu_defn}) that confirms once again that the quasiparticles exist even on a wave packet of a single neutron. The nonlocal operators $1/(p_ck)$ and $1/(p_ck')$ appearing in $\pi^{\mu\nu}$ have to be defined precisely for the effective Maxwell equations \eqref{Max_eqns_eff_IR} to be causal. It is not difficult to verify that the causal Maxwell equations \eqref{Max_eqns_eff_IR} follow from the action functional
\begin{equation}\label{action_IR}
    S[A_\mu(x)]=-\frac{1}{4}\int d^4x F_{\mu\nu}(x)F^{\mu\nu}(x) -\int d\tau m_\s(\tau) H^\s(x(\tau)),\qquad H^\s=\frac12\e^{\s\mu\rho\la} F_{\mu\rho} \dot{x}_\la,
\end{equation}
where $H^\s$ is the magnetic field strength vector, $\dot{x}^\mu=p^\mu_c/M$, and $m_\s(\tau)$ obeys the equations
\begin{equation}\label{magn_mom_evol}
    \frac{dm_\s(\tau)}{d\tau}=\mu^2_p N_h\big[F_{\s\nu}(x(\tau)) -\dot{x}_\s \dot{x}^\la F_{\la\nu}(x(\tau)) \big] s_h^\nu.
\end{equation}
The causal solution to this equation should be substituted into the equations resulting from a variation of the action \eqref{action_IR}. In the rest frame of a neutron, equation \eqref{magn_mom_evol} becomes
\begin{equation}
    \dot{\mathbf{m}}=\mu^2_p N_h[\mathbf{s}_h,\mathbf{H}].
\end{equation}
The second term in the action \eqref{action_IR} describes the standard interaction of the magnetic moment $m_\s$ with the external magnetic field. Equation \eqref{magn_mom_evol} governs the evolution of the magnetic moment induced by the external electromagnetic field (see, e.g., Sec. 45 of \cite{LandLifshCTF.2} and Sec. 69 of \cite{LandLifshST2}). The total magnetic moment has the form
\begin{equation}
    \mu_p N_h s^\s_h+m^\s,
\end{equation}
where the second term is assumed to be small in comparison with the first one. In other words, in the infrared limit of a coherent Compton process, a spin polarized neutron behaves as a neutral point particle with dynamical magnetic moment, the dynamics of the magnetic moment being determined by equation \eqref{magn_mom_evol} in the leading order of perturbation theory. The additional degrees of freedom existing on the neutron wave packet and discussed thoroughly in the previous section are reduced to the dynamical magnetic moment.

\section{Conclusion}

Let us sum up the results. We have considered the Compton process for neutrons where the initial states of both the photons and the neutrons are described by some density matrices of a general form and the final state of neutrons is not recorded. The energies of photons have been assumed much lower than the $\pi$-meson rest energy in the neutron rest frame. We have investigated only the coherent contribution to the scattering amplitude \cite{pra103,KazSol2022,KazSol2023,KRS2023,radet,AKS2025} that describes the interference of the state of free passed photons with its scattered part caused by neutrons in Compton scattering. Thereby, we have studied the hologram of the neutron state. As follows from the general theory \cite{KazSol2022,KazSol2023,AKS2025}, the amplitude of coherent Compton scattering is the same as if the photons were scattered by the medium with a certain dielectric susceptibility tensor even in the case of scattering by a singe neutron. We have derived the explicit expression for this tensor both on- and off-shell. The off-shell expression for the susceptibility tensor of a gas of neutrons or a single neutron wave packet is, up to a common factor, the photon polarization operator in the presence of neutrons. We have obtained the simple expressions for this polarization operator in the short wavelength approximation \eqref{polariz_oper_shrt_wl} and in the infrared limit \eqref{pi_munu_defn}, \eqref{Max_eqns_eff_IR}.

In investigating coherent Compton scattering, we have found the general expression \eqref{rho2_ph} for the one-particle density matrix of scattered photons. We have scrutinized this density matrix in the small recoil limit in the case of nonrelativistic neutrons and have deduced the explicit expressions for a change of the Stokes parameters \eqref{non_pol_dens_out}, \eqref{pol_dens_out}, \eqref{half_pol_dens_out}. In the particular case of a homogeneously spin polarized one-particle density matrix of incident photons, we have deduced the evolution equation for the Stokes vector \eqref{de_xi} of scattered photons. It has been shown that, in multiple coherent scattering by neutrons prepared in the same state, the scattered photon tends to be circularly polarized whereas the plane of its linear polarization is rotated. We have particularized the general formulas \eqref{non_pol_dens_out}, \eqref{pol_dens_out}, \eqref{half_pol_dens_out} for a variation of the Stokes parameters to the case of Gaussian initial one-particle density matrices of photons and neutrons and have derived the explicit expressions \eqref{pol_dens_out_Gauss}, \eqref{non_pol_dens_out_Gauss}, \eqref{half_pol_dens_out_Gauss} for this variation corroborating the conclusions drawn from the general formulas. The plots of the parameters specifying a variation of the Stokes vector are presented in Figs. \ref{kappa_polarized_plots}, \ref{kappa_unpolarized_plots}. Notice that the magnitude of the effect can be enhanced by the Lorentz factor of neutrons when the photons are scattered by the ultrarelativistic neutrons moving in almost the same direction as the photons. This conclusion follows directly from the general formula \eqref{rho2_ph} with account for the relation \eqref{de-func_energy} and can be easily understood without any calculations as the interaction time of photons with neutrons increases in that case.

We have obtained the general expression for the dielectric susceptibility tensor of a neutron wave packet and a neutron gas \eqref{chi_gen}. We have considered the unpolarized and polarized neutron states and the relativistic and non-relativistic cases, have found the explicit expressions \eqref{chi_xi_non0}, \eqref{chi_xi0}, \eqref{chi_xi00} for the susceptibility tensor of neutrons in these cases, and have provided the physical interpretation to them. In particular, we have revealed that, in coherent Compton scattering, the spin polarized neutrons behave as a gyrotropic medium. As for the unpolarized neutrons, their susceptibility tensor has the form that arises in describing dynamo-optical phenomena in inhomogeneously moving fluids \cite{LandLifshECM}.

We have also investigated the off-shell expression for the photon polarization operator in the presence of a neutron or a gas of neutrons. In the short wavelength approximation, it turns out that this polarization operator possesses pole singularities. As is known (see, e.g., \cite{WeinbergB.12}), such singularities signalizes the presence of additional degrees of freedom in the theory -- quasiparticles. Such quasiparticles exist even on the wave packet of a single neutron, have the same dispersion law as in the case of a single electron wave packet \cite{AKS2025} but with the replacement of the electron mass by the neutron mass, and can be called plasmons. Hybridization of these degrees of freedom with the electromagnetic field gives rise to plasmon-polaritons on a wave packet of a single neutron or in a neutron gas. There are eight independent plasmon-polariton modes. We have derived the explicit expressions for these modes \eqref{plasmon_polariton_modes} and their dispersion laws \eqref{plasmon_polariton_disp_law} for both the unpolarized and polarized neutron states that are narrow in the momentum space. Quite surprisingly, there are the longitudinal plasmon-polariton modes describing oscillations of a pure electric field in the neutron rest frame. Moreover, some of the branches of the plasmon-polariton dispersion law prove to be tachyonic and unstable in certain domains of plasmon-polariton momenta (see Fig. \ref{dispersion_laws_plots}). We have shown that this instability describes a spontaneous creation of the magnetic field and so a neutron state that is sufficiently narrow in the momentum space is ferromagnetic. We have found the estimate \eqref{instab_region_est3} for the parameters of the one-particle density matrix of neutrons when the ferromagnetic instability is realized. This estimate shows that, for sufficiently high densities and at low temperatures, a nondegenerate gas of neutrons is spontaneously magnetized. In particular, the nondegenerate gas of ultracold neutrons \cite{Ignatovich1996,Serebrov2011,Pokotilovski2018,Lauss2021} at a temperature of $1.06$ neV and with the mass density of order of one half of the water mass density at normal conditions is ferromagnetic. Unfortunately, such densities of an ultracold neutron gas are unattainable at present in laboratories. Nevertheless, one can speculate that this mechanism for generation of strong magnetic fields can be realized in creating the magnetic fields near neutron stars in addition to the other ways of generation of such fields \cite{HaensPtoYak2007,Ginzburg1964,Urpin1986,Thompson1993,Igoshev2025,Price2006,Potekhin2010,Dong2013}. As follows from the estimate \eqref{instab_region_est3}, the critical density is only of order $2\times10^{-4}$ of the nuclear density at an effective temperature of $10^6$ K and grows linearly with temperature.

We have obtained the infrared limit of the off-shell polarization operator of a photon in the presence of spin polarized neutrons. We have shown that, as expected, in the infrared limit of coherent Compton scattering, the neutron behaves as a point particle with dynamical magnetic moment whose dynamical part obeys equation \eqref{magn_mom_evol} in the leading order of perturbation theory. The additional degrees of freedom manifesting explicitly in the short wavelength approximation boil down to the dynamical part of the magnetic moment. In that case, we have deduced the effective Maxwell equations \eqref{Max_eqns_eff_IR} and the effective action functional \eqref{action_IR} with constraint \eqref{magn_mom_evol} for them.

The theoretical results that we have obtained in the present paper can be verified experimentally in the processes of coherent interaction of electromagnetic fields with neutrons. Despite the fact that the effects are rather small they are detectable even with the present experimental facilities (see Fig. \ref{kappa_polarized_plots}) as there are sources of neutrons with $10^{14}$ neutrons per bunch and larger \cite{Henderson2014,Anderson2016}. Some of the results can be relevant for physics of ultracold neutrons and neutron stars. However, we postpone these possible applications for a further investigation. The tachyonic tails of the plasmon-polariton dispersion laws at large momenta \eqref{disp_law_asympt} also deserve a further exploration for the concrete profiles of one-particle density matrices of polarized neutrons. Also note that some of the results of the paper can be applied to coherent scattering of photons by a dilute gas of identical neutral particles with spin $1/2$ and nonzero magnetic moment. Note however that for a straightforward application of the results of the paper the wavelengths of scattered photons must be much larger than the size of these molecules and all the molecules should be identical, in particular, they should be in the same internal state. As for the instability described in Sec. \ref{Off-shell_Polarization_Operator}, the conditions for its realization, apparently, cannot be realized even for neutral atoms. The atoms at such large densities and small temperatures as in \eqref{instab_region_est3} will form bound states such as molecules, molecular conglomerates, a liquid, or a solid.

\paragraph{Acknowledgments.}
This research was supported by the TPU development program Priority 2030.


\appendix
\section{Tensors $M_{\mu\nu}$ and $N_{a\mu\nu}$}\label{M_and_N_App}

The tensors $M_{\mu\nu}$ and $N_{a\mu\nu}$ introduced in \eqref{polar_vec_prods} are Hermitian, transverse to the corresponding momenta, and spacelike:
\begin{equation}
\begin{aligned}
	M^*_{\mu\nu}(k,k')&=M_{\nu\mu}(k',k), &\qquad N^*_{a\mu\nu}(k,k')&=N_{a\nu\mu}(k',k),\\
	k^\mu M_{\mu\nu}(k,k')&= M_{\mu\nu}(k,k')k'^\nu=0,&\qquad k^\mu N_{a\mu\nu}(k,k')&= N_{a\mu\nu}(k,k')k'^\nu=0,\\
    t^\mu M_{\mu\nu}(k,k')&= M_{\mu\nu}(k,k')t^\nu=0,&\qquad t^\mu N_{a\mu\nu}(k,k')&= N_{a\mu\nu}(k,k')t^\nu=0.
\end{aligned}
\end{equation}
They possess the following properties
\begin{equation}
\begin{gathered}
    M^\mu_{\ \rho}(k_1,k_2) M^\rho_{\ \nu}(k_2,k_3)=-M^\mu_{\ \nu}(k_1,k_3),\\
    M_\mu^{\ \rho}(k_1,k_2) N_{a\rho\nu}(k_2,k_3)=-N_{a\mu\nu}(k_1,k_3),\qquad N_{a\mu\rho}(k_1,k_2) M^{\rho}_{\ \nu}(k_2,k_3)=-N_{a\mu\nu}(k_1,k_3),\\
    N_{a\mu\rho}(k_1,k_2) N^{\ \rho}_{a\ \nu}(k_2,k_3)=\de_{ab}M_{\mu\nu}(k_1,k_3) -i\e_{abc} N_{c\mu\nu}(k_1,k_3),\\
    M^\mu_{\ \mu}(k,k)=-2,\qquad N^{\ \mu}_{a\ \mu}(k,k)=0,\\
    M_{\mu\nu}(k_1,k_2) M_{\rho\s}(k_3,k_4) +N_{a\mu\nu}(k_1,k_2) N_{a\rho\s}(k_3,k_4)=2 M_{\mu\s}(k_1,k_4) M_{\rho\nu}(k_3,k_2).
\end{gathered}
\end{equation}
The above properties imply that $-M^\mu_{\ \nu}(k,k)$ is a Hermitian projector onto the two-dimensional linear subspace orthogonal to $k^\mu$ and $t^\mu$. Therefore,
\begin{equation}
    -M_{\mu\nu}(k,k)=\eta_{\mu\nu} -\frac{k_{(\mu} t_{\nu)}}{(kt)} +\frac{k_\mu k_\nu}{(kt)^2},
\end{equation}
where $k^2=0$ and the parentheses around a pair of indices denote a symmetrization without the factor $1/2$.

If the polarization vectors are chosen as in \eqref{polarization_vects}, then we obtain the explicit expressions
\begin{equation}
	M^{\mu\nu}(k,k')=M^\mu_{\ \mu'}(k,k)\frac{m^{\mu'\nu'}(k,k')}{k_\perp k'_\perp}M_{\nu'}^{\ \nu}(k',k'),
    \qquad N^{\mu\nu}_{a}(k,k')=M^\mu_{\ \mu'}(k,k)\frac{n^{\mu'\nu'}_{a}(k,k')}{k_\perp k'_\perp}M_{\nu'}^{\ \nu}(k',k'),
\end{equation}
where
\begin{equation}
\begin{split}
	m^{\mu\nu}(k,k')&=\eta^{\mu\nu}\big[(kk')-(kt)(k't)+(kd)(k'd)\big] - k'^\mu k^\nu
    - d^\mu k^\nu(k'd) -k'^{\mu}d^\nu(kd) +d^\mu d^\nu (kk'),\\
	n_1^{\mu\nu}(k,k')&= d^\mu\e^{\nu\rho\la\s}t_\rho k'_\la d_\s (kt) +d^\nu\e^{\mu\rho\la\s}t_\rho k_\la d_\s (k't),\\
	n_2^{\mu\nu}(k,k')&=-id^\mu\e^{\nu\rho\la\s}t_\rho k'_\la d_\s (kt) +id^\nu\e^{\mu\rho\la\s}t_\rho k_\la d_\s (k't),\\
	n_3^{\mu\nu}(k,k')&=m^{\mu\nu}(k,k') -2 d^\mu d^\nu (kt)(k't).
\end{split}
\end{equation}
In the small recoil limit, using the representation \eqref{pc_kc_q}, we have
\begin{equation}
\begin{split}
	M^{\mu\nu}(k,k')&=M^\mu_{\ \mu'}(k,k)\Big[-\eta^{\mu'\nu'}+\frac{(k_cd)}{k_{c\perp}^{2}} d^{[\mu'} q^{\nu']}+O(q^2)\Big]
    M_{\nu'}^{\ \nu}(k',k'),\\
	N_1^{\mu\nu}(k,k')&=M^\mu_{\ \mu'}(k,k)\Big[\frac{(k_ct) }{k_{c\perp}^2} d^{(\mu'}\e^{\nu')\rho\la\s}t_\rho k^c_\la d_\s +O(q)\Big]
    M_{\nu'}^{\ \nu}(k',k'),\\
	N_2^{\mu\nu}(k,k')&=M^\mu_{\ \mu'}(k,k)\Big[-i\frac{(k_ct) }{k_{c\perp}^2} d^{[\mu'}\e^{\nu']\rho\la\s}t_\rho k^c_\la d_\s +O(q)\Big]
    M_{\nu'}^{\ \nu}(k',k'),\\
    N_3^{\mu\nu}(k,k')&=M^\mu_{\ \mu'}(k,k)\Big[-\eta^{\mu'\nu'}-2\frac{(k_ct)^2}{k_{c\perp}^2}d^{\mu'}d^{\nu'}+O(q)\Big]
    M_{\nu'}^{\ \nu}(k',k'),
\end{split}
\end{equation}
where square brackets at a pair of indices denote an antisymmetrization without the factor $1/2$.

\section{Contractions of the tensors $M_{\mu\nu}$ and $N_{a\mu\nu}$ with the tensors  $G_3^{\mu\nu}$ and $Z_3^{i\mu\nu}$}\label{Contractions_App}

In the small recoil approximation, $q^\mu\rightarrow0$, and in the non-relativistic limit, $|\spp_c|/p_0\ll1$, the contractions of the tensor $Z_3^{i\mu\nu}$ defined in \eqref{G_n_Z_n_defn} with the tensors $M_{\mu\nu}$, $N_{a\mu\nu}$ take the form
\begin{equation}\label{Z_contractions}
\begin{split}
	M_{\nu\mu}(k,k')Z_3^{i\mu\nu}&= -\frac{2i}{k_c^0}\Big(\e^{0\mu\nu i}k^c_\mu q_\nu -2\frac{n_c^i (\spn_c\mathbf{d})
    \e^{0\mu\nu\rho}d_\mu k^c_\nu q_\rho}{n^{2}_{c\perp}}\Big)+O(q^3),\\
	N_{1\nu\mu}(k,k')Z_3^{i\mu\nu}&=-2i \Big(q^i-2\frac{(\spq\mathbf{d})
    [d^i-n_c^i(\spn_c\mathbf{d})]}{n^{2}_{c\perp}}\Big)+O(q^3),\\
	N_{2\nu\mu}(k,k')Z_3^{i\mu\nu}&=4k_c^i-4p_c^i \frac{k_c^0+(k_cp_c)/M}{p_c^0+M}+O(q^2),\\
	N_{3\nu\mu}(k,k')Z_3^{i\mu\nu}&=\frac{2i}{k_c^0} \Big(\e^{0\mu\nu i}k^c_\mu q_\nu+2\frac{(\spq\spd)
    \e^{0\mu\nu i}d_\mu k^c_\nu}{n_{c\perp}^2}\Big)+O(q^3),
\end{split}
\end{equation}
where it has been assumed that $t^\mu=(1,0,0,0)$. The corresponding contractions for the tensor $G_3^{\mu\nu}$ defined in \eqref{G_n_Z_n_defn} are written as
\begin{equation}\label{G_contractions}
\begin{split}
	M_{\nu\mu}(k,k')G_3^{\mu\nu}&= \frac{q^2}{M}+O(q^4),\\
	N_{1\nu\mu}(k,k')G_3^{\mu\nu}&= -2\frac{(\spq\spd)\e^{0\mu\nu\rho}d_\mu n^c_\nu q_\rho}{M n_{c\perp}^2}+O(q^4),\\
	N_{2\nu\mu}(k,k')G_3^{\mu\nu}&=\frac{2i}{M} \frac{\e^{0\mu\nu\rho} k^c_\mu p^c_\nu q_\rho}{p_c^0+M}+O(q^3),\\
	N_{3\nu\mu}(k,k')G_3^{\mu\nu}&= -\frac{q^2}{M}-2\frac{(\spq\spd)^2}{M n_{c\perp}^2} +O(q^4).
\end{split}
\end{equation}
Notice that only the small recoil approximation has been used in the expressions for contractions with the tensor $N_{2\mu\nu}$, i.e., these expressions are also valid in the relativistic case. It should be noted that the order of contractions \eqref{Z_contractions}, \eqref{G_contractions} with respect to powers of $q^\mu$ in the relativistic case is the same as in the non-relativistic limit considered above. As is seen, the contractions with the tensor $N_{2\mu\nu}$ yield the leading contributions in the small recoil limit.

\section{Tensors $G_3^{\mu\nu}$ and $Z_3^{i\mu\nu}$ in the small recoil limit}

In the small recoil limit, $q^\mu\rightarrow0$, the contractions of the tensors $G_3^{\mu\nu}(\spp,\spp')$ and $Z_3^{i\mu\nu}(\spp,\spp')$ with the polarization vectors ($e_\mu=e^{(\la)}_\mu(\spk)$, $e'_\mu=e^{(\la')}_\mu(\spk')$) are written as
\begin{equation}\label{ee_G3_Z3}
\begin{split}
    \bar{e}'_\mu e_\nu G^{\mu\nu}_3&= -\frac{(\bar{e}'e)q^2}{M+p_c^0}\Big(1+\frac{k_c^0M}{(k_cp_c)}\Big) +\frac{(\bar{e}'q)(ep_c)}{M+p_c^0} \frac{q_0-2k_c^0}{2M} +\frac{(\bar{e}'p_c)(eq)}{M+p_c^0} \frac{q_0+2k_c^0}{2M} +\frac{(\bar{e}'p_c)(ep_c)k_c^0q^2}{M(M+p_c^0)(k_cp_c)} +\\
	&+\frac{(\bar{e}'q)(eq)}{M+p_c^0} \Big(1-\frac{p_c^0}{M}+\frac{2k_0M}{(k_cp_c)}\Big)+o(q^2),\\
	\bar{e}'_\mu e_\nu  Z^{i\mu\nu}_3&=-\frac{2i\e^{\mu\nu\rho\la}\bar{e}'_\mu e_\nu
    k^c_\rho}{M(M+p_c^0)} \big(p^c_\lambda+M\delta^0_\lambda\big)p_c^i -2i\e^{i\mu\nu\rho} \bar{e}'_\mu e_\nu k^c_\rho +o(1),\\
\end{split}
\end{equation}
In the off-shell case, we have from \eqref{G_n_Z_n_defn} in the small recoil limit that
\begin{equation}\label{G3_Z3_off-shell}
\begin{split}
	G_3^{\mu\nu}&=-\frac{M k_c^4}{(k_cp_c)^2-k_c^4/4}\Big[\eta^{\mu\nu}
    -\frac{p_c^\mu p_c^\nu}{M^2}+\frac{k_c^{(\mu}p_c^{\nu)}(k_cp_c)}{M^2k_c^2}-\frac{k_c^\mu k_c^\nu}{k_c^2}\Big(1+\frac{(k_cp_c)^2}{M^2k_c^2}\Big)\Big]+o(1),\\
	\xi_i Z_3^{i\mu\nu}&=\frac{2i}{(k_cp_c)^2-k_c^4/4}\Big[(k_cp_c)(k_cs^h)
    \e^{\mu\nu\rho\sigma}k^c_\rho p^c_\sigma-\frac{k_c^4}{4}\e^{\mu\nu\rho\s}k^c_\rho s^h_\s\Big]+o(1),
\end{split}
\end{equation}
where the neutron spin $4$-vector $s^\mu_h$ was introduced in \eqref{s_h_mu_defn}.


\end{document}